\documentclass[aoas,MSNbibl,nameyear,seceqn,dvips]{arximspdf}
\usepackage{dcolumn}
\usepackage{graphicx}

\doi{10.1214/12-AOAS608} 
\volume{7}
\issue{1}
\pubyear{2013}
\firstpage{126}
\lastpage{153}

\makeatletter
\def\mid{\vert}
\newcolumntype{d}[1]{D{.}{.}{#1}}

\newtheorem{theorem}{Theorem}
\newtheorem{lemma}[theorem]{Lemma}
\makeatother

\begin{document}
\begin{frontmatter}

\title{Bayesian analysis of dynamic item response models in educational
testing\thanksref{T1}}
\thankstext{T1}{Supported by NSF Grant DMS-10-07773 and
by MetaMetrics Inc.}
\runtitle{Dynamic item response models}

\begin{aug}
\author[A]{\fnms{Xiaojing} \snm{Wang}\ead[label=e1]{xiaojing.wang@uconn.edu}},
\author[B]{\fnms{James O.} \snm{Berger}\corref{}\ead[label=e2]{berger@stat.duke.edu}}
\and
\author[C]{\fnms{Donald S.} \snm{Burdick}\ead[label=e3]{dburdick@lexile.com}}
\runauthor{X. Wang, J. O. Berger and D. S. Burdick}
\affiliation{University of Connecticut, Duke University and
MetaMetrics, Inc.}
\address[A]{X. Wang\\
Department of Statistics\\
University of Connecticut\\
215 Glenbrook Road, U-4120\\
Storrs, Connecticut 06269-4120\\
USA\\
\printead{e1}}

\address[B]{J. O. Berger\\
Department of Statistical Science\\
Duke University, P.O. Box 90251\\
Durham, North Carolina 27708-0251\\
USA\\
\printead{e2}}

\address[C]{D. S. Burdick\\
MetaMetrics, Inc., Suite 120\\
1000 Park Forty Plaza Drive\\
Durham, North Carolina 27713\\
USA\\
\printead{e3}}
\end{aug}

\received{\smonth{6} \syear{2011}}
\revised{\smonth{10} \syear{2012}}

%
\begin{abstract}
Item response theory (IRT) models have been widely used in
educational measurement testing. When there are repeated
observations available for individuals through time, a dynamic
structure for the latent trait of ability needs to be incorporated
into the model, to accommodate changes in ability. Other
complications that often arise in such settings include a violation
of the common assumption that test results are conditionally
independent, given ability and item difficulty, and that test item
difficulties may be partially specified, but subject to uncertainty.
Focusing on time series dichotomous response data, a new class of
state space models, called Dynamic Item Response (DIR) models, is
proposed. The models can be applied either retrospectively to the
full data or on-line, in cases where real-time prediction is needed.
The models are studied through simulated examples and applied to a
large collection of reading test data obtained from MetaMetrics,
Inc.
\end{abstract}

%

\end{frontmatter}

\section{Introduction}\label{sec1}

\subsection{Background}\label{sec1.1}
Item response theory (IRT) models are frequently used in
modeling dichotomous data from educational
tests, since they allow separate assessment of the ability of examinees
and effectiveness of the test items.
A typical one-parameter IRT model is of the form
%
\begin{equation}\label{it-11}
\operatorname{ Pr}(X_{il}=1 \mid\theta_i,d_l)=\mathrm{ F}
(\theta_i-d_l),
\end{equation}
where $\theta_i$ indicates the ability of the $i$th person; $d_l$
indicates the difficulty of the $l$th test item; the item response
variable $X_{il}$ could be either 0 or 1, corresponding to whether
the $l$th test item taken by the $i$th person is answered
correctly or not; and the item characteristic curve, $\mathrm{
F(\cdot)}$, is a cumulative distribution function (c.d.f.) from a
continuous distribution. When $\mathrm{ F(\cdot)}$ is the standard
logistic c.d.f., the one-parameter IRT model (\ref{it-11}) becomes the
famous Rasch model
%
\begin{equation}\label{it-12}
\operatorname{ Pr}(X_{il}=1 \mid\theta_i,d_l)=
\frac{\exp(\theta_i-d_l)}{1+\exp(\theta_i-d_l)}.
\end{equation}
If $\mathrm{ F}(\cdot)=\Phi(\cdot)$, where $\Phi(\cdot)$ is the standard
normal c.d.f., then
%
\begin{equation}\label{it-13}
\operatorname{ Pr}(X_{il}=1 \mid\theta_i,d_l)=\Phi(
\theta_i-d_l)
\end{equation}
defines the one-parameter Normal Ogive or Probit model. We will
focus on the former model in the paper, for reasons to be discussed
later, although analysis of the Probit model is actually easier and
can be done with a simplified version of the methodology developed
here.

The development of item response theory from the classical point of
view owes much to the pioneering work of \citet{Lord}, Rasche (\citeyear{Rasch})
and their colleagues. Among the many noteworthy contributions are
\citet{Andersen} and \citet{Bock}.

In classical IRT, it is assumed that the $X_{il}$ are independent,
given the person's ability $\theta_i$ and the difficulty levels
$d_l$. This is often referred to as the \textit{local independence}
assumption. There are situations in which this assumption is
violated. One such is computer adaptive testing, wherein the
selection of the next test item typically depends specifically on
the previous questions and answers.

The situation is less clear with what is studied herein,
MetaMetrics' educational assessment program called Computer Adaptive
Instruction and Testing (CAIT). With CAIT, a test pool of articles
is selected for the student based on an estimate of his/her current
ability; the student selects an article from this pool and the test
questions (described later) are then generated before reading
commences. Thus, in the environment of the CAIT, the possible
violation in the local independence would arise from sources such as
article selection by the student and test questions related to the
same article so that overall understanding of the article could
affect all answers; in this paper, such possible effects will be
called \textit{test effects}. Other factors that could cause violation
of the local independence include health status and emotional status
of the student on a given day; these will be referred to as {\em
daily effects}. In the MetaMetrics scenario, there had been no
previous demonstration of the violation of the local independence
through the presence of test effects or daily effects, and there was
a considerable interest in establishing such presence for possible
enhancement of current models.

Pioneering papers that addressed the local dependence were Stout
(\citeyear{StoutA}, \citeyear{Stout}), who introduced the essential dimensionality and the
essential independence of a collection of test items, and \citet{Gibbons},
who considered the conditional dependence within
identified subsets of items by allowing random effects in the
analysis. More recent work in this direction is testlet response
theory modeling, proposed by \citet{Bradlow}. They
defined the testlet as a subset of items; for example, they defined
a reading comprehensive section in the SAT as the testlet. They then
modified the classic IRT models by including a random effect term to
represent the common factor affecting the responses in the testlet.
Another approach to handle the local dependence is by the
introduction of Markov structure, such as \citet{Jannarone} where the
conjunctive IRT kernel was introduced. A more recent paper concerned
is \citet{Andrich}, where they modified the Rasch model
by allowing the conditional probability of a response to an item to
depend on the answer of a previous item.

For the modeling in this paper, the random effect approach will be
followed. Indeed, two levels of random effects will be introduced
to model the daily effects and test effects, respectively.

Another essential generalization of the IRT model lies in their
applicability to analyze longitudinal data, that is, to deal with
scenarios in which an individual is tested repeatedly over time;
then, the interest typically centers on the growth of an ability of
the individual. Embretson (\citeyear{Emberetson}) and Marvelde et al. (\citeyear{Marvelde})
presented a multidimensional Rasch model to represent the change of
an ability as an initial ability and one or more modifiabilities.
Based on the belief that a person's ability growth would be
increasing over time, \citet{Albers}, \citet{Tan} and
Johnson and Raudenbush (\citeyear{Johnson}) used linear or polynomial regression
of the time variable to measure the growth of an ability; their
analysis required the same time span and testing points for all
examinees. \citet{Martin} modeled the transition of a
voting preference as a first-order Markov process, where they
assumed voting preference changes from the previous time point to a
new point by a random shock; this work did not incorporate a time
trend. \citet{Park} supposed that changes in a voting preference were
subject to discrete agent-specific regime changes and modeled the
indicator of the preference regime changes as a first-order Markov
process. Bartolucci, Pennoni and
Vittadini (\citeyear{Bartolucci}) analyzed test scores in
mathematics observed over 3 years for public and private middle
school students by a multilevel latent Markov Rasch model, where
they described the dynamic transition of different levels of the
individual ability also via a first-order Markov process.

Our approach to the longitudinal issue is based on a new class of
dynamic linear models (DLM's) [see West and Harrison (\citeyear{West}) for
background on DLM's]. The literature on DLM's or state space models,
in the framework considered here of longitudinal binomial data,
includes, for example, \citet{Carlin}, \citet{Fahrmeir} and
Czado and Song (\citeyear{Czado}) and the last three papers mentioned in the
previous paragraph. Our models are distinguished from the literature
by simultaneously allowing for the following features: (i)
observations at variable and irregular time points; (ii)~continuously changing ability, but with incorporation of knowledge
concerning trends (e.g., increasing ability over time) in a
nondogmatic way (thus accommodating, say, a drop in reading ability
over a summer vacation); (iii)~an analysis that is either individual
or hierarchical across a group of individuals, the latter allowing
for ``borrowing strength'' in estimates of certain overall
parameters; (iv)~either a retrospective analysis based on the full
data or a real-time analysis and prediction for an individual based
on the data to date.

Moreover, we consider the case in which the test item difficulties
are nominally specified, as in CAIT, where the test items are often
computer-generated and have theoretically determined difficulties.
The actual item difficulties are quite uncertain, however, this
uncertainty is also accommodated in our analysis. Previous papers
that introduced random effects for item parameters include Sinharay,
Johnson and Williamson (\citeyear{Sinharay}) and DeBoeck (\citeyear{Boeck}).

\subsection{Testbed application}\label{sec1.2}
The model developed in this paper is motivated by CAIT testing, as
developed by MetaMetrics Inc. The main applied goals are as follows:
\begin{itemize}
\item The original
goal is to assess the appropriateness of the local independence
assumption for this type of data. This evolves into the goal of
better understanding the nature of the daily and test effects.
\item A second goal is to understand the growth in ability of
students, by retrospectively producing the estimated growth
trajectories of their latent abilities in the study.
\item A third goal is to enable on-line prediction of one's ability (based
solely on data obtained up to that point), to enable a better
assignment of reading materials to match his/her ability and to
enable teachers to better assist students.
\end{itemize}

The data considered is from a school district in Mississippi and
consisted of 1983 students who registered over two years in a CAIT
reading test program conducted by MetaMetrics Inc. The students
were in different grades and entered and left the program at
different times between 2007 and 2009. Individuals took tests on
different days and had different time lapses between tests. Because
of the long periods of testing, a fully adaptive model accommodating
continual changes in ability is needed.

The data was generated during sessions in which a student read an
article selected from a large bank of available articles. The
articles in this bank had been assigned text complexity measured in
Lexiles, using the Lexile Receptive Analyzer $\circledR$, a software
developed by MetaMetrics Inc. to evaluate the semantic and syntactic
complexity of a text. The Lexile measure represents either an
individual's reading ability or the complexity of a text. The scale
for Lexiles ranges from 0 to 1800, with 0 indicating no reading
ability and 1800 being the maximum.

A session begins like this: a student selects from a generated list
of articles having Lexile complexities in a range targeted to the
current estimate of the student's ability. For the selected article,
a subset of words from the article are eligible to be \textit{clozed},
that is, removed and replaced by a blank. The computer, following a
prescribed protocol, randomly selects a sample of the eligible words
to be clozed and presents the article to the student with these
words clozed. When a blank is encountered while reading the
article, the student clicks it and then the true removed word along
with three incorrect options called foils is presented. As with the
target word, the foils are selected randomly according to a
prescribed protocol. The student selects a word to fill in the blank
from the four choices and an immediate feedback is provided in the
form of the correct answer.

The dichotomous items produced by this procedure are called
``Auto-Generated-Cloze'' items. They are single-use items generated at
the time of an encounter between a student and an article. If
another student selects that same article to read, a new set of
target words and foils is selected. Although it is not strictly
impossible for an individual item to be taken by more than one
student, such an occurrence is highly improbable. As a consequence,
it is not feasible to obtain data-based estimates of item
calibration parameters.

Instead, the difficulties of the items generated for an encounter
between a student and an article can be modeled as a sample from an
ensemble of item difficulties associated with the article. The text
complexity in Lexiles provides a theoretical value for the ensemble
mean. An estimated student ability in combination with assumptions
about the ensemble allows calculation of a predicted success rate
for the encounter. A comparison of the observed success rate with
predicted, aggregated over many encounters, provides a basis for
assessing the viability of the assumptions incorporated into the
model. The predicted success rates in Table 1 in \citet{Stenner}
include the assumption that the mean of the ensemble of item
difficulties for an article is given by its theoretical text
complexity. The agreement with observed success rates supports that
assumption.

Although MetaMetrics data is typically presented in Lexile units,
there is a simple linear transformation from Lexiles to logit units.
We will utilize the more common logit units for all data and
results in this paper. Note that this also motivates the use of
the logistic IRT model in this paper---to preserve compatibility
with the MetaMetrics data.

\subsection{Preview}\label{sec1.3}
Because of the complexity of the model considered (and of the
testbed data set), as well as the need to incorporate prior
information into the model, the analysis will be carried out using
Bayesian methodology and Markov chain Monte Carlo (MCMC)
computational techniques. A side benefit of using these
methodologies is that all uncertainties in all quantities are
combined in the overall assessment of inferential uncertainty. The
MCMC procedure utilizes a novel combination of Gibbs sampling
together with a block sampling scheme involving forward filtering
and backward sampling.

In Section~\ref{sec2} we formally describe the proposed models to capture
the dynamic changes in a person's ability as well as the local
dependence between item responses. Section~\ref{sec3} presents the MCMC
strategy to carry out the statistical inference. Section~\ref{sec4} tests the
methodology on some simulated examples (where the truth is known).
Section~\ref{sec5} applies the proposed models to the MetaMetrics data set.
Section~\ref{sec6} draws conclusions from both statistical and psychological
sides, and points out some directions for future studies.

\section{Dynamic item response (DIR) models}\label{sec2}
This section formally introduces the proposed one-parameter DIR
model. Although the focus is on generalizing one-parameter IRT
models, it would be straightforward to similarly generalize
two-parameter or three-parameter IRT models.

\subsection{The observation equation in DIR models}\label{sec2.1}
In a typical one-parameter IRT model (\ref{it-11}), the index of
the item response $X_{il}$ indicates the correctness of the $i$th
person's answer to the $l$th question in a single test. Consider
the more involved situation in which the individual completes a
series of tests within a given day and over different days. Thus,
the item response variable is $X_{i,t,s,l}$, which corresponds to
the correctness of the answer of the $l$th item in the $s$th test
on the $t$th day taken by the $i$th person. Here, $i=1,\ldots,n$;
$t=1,\ldots,T_i$; $s=1,\ldots,S_{i,t}$; and $l=1,\ldots,K_{i,t,s}$.

Likewise, let $d_{i,t,s,l}$ represent the difficulty level of the
$l$th item in the $s$th test at the $t$th day taken by the $i$th
person. As described in the \hyperref[sec1]{Introduction}, we
model the test difficulties as being nominally specified, but with
uncertainty. Thus, we write
%
\begin{equation} \label{irt-11}
d_{i,t,s,l}=a_{i,t,s}+\varepsilon_{i,t,s,l},
\end{equation}
where $a_{i,t,s}$ indicates the ensemble mean difficulty for the
items in the $s$th test taken by the $i$th person on the $t$th
day, and $\varepsilon_{i,t,s,l}$ is the random deviation from this
ensemble mean difficulty for the $l$th item within the $s$th test.
In the scenario we consider, the value of $a_{i,t,s}$ is assumed to
be known, from the theoretical analysis of text complexity, while it
is assumed that $\varepsilon_{i,t,s,l}$ is a normal distribution with
zero mean and specified variance $\sigma^2$ from the test design in
the CAIT testing, which is denoted as $\varepsilon_{i,t,s,l}\sim
\mathcal{N}(0,\sigma^2)$.

As mentioned in the \hyperref[sec1]{Introduction}, we will also incorporate a term of
daily random effects, $\varphi_{i,t}$, as well as a term of test
random effects, $\eta_{i,t,s}$, to account for the possible local
dependence factors when person $i$ takes several tests during day
$t$. It is assumed that $\varphi_{i,t}\sim
\mathcal{N}(0,\delta_{i}^{-1})$ and, letting
${\eta}_{i,t}=(\eta_{i,t,1},\ldots,\eta_{i,t,S_{i,t}})'$ denote the
vector of test random effects on day $t$ for individual $i$, that
${\eta}_{i,t} \sim\mathcal{N}_{S_{i,t}}(0,\tau_i^{-1}\mathbf{I}
\mid\sum_{s=1}^{S_{i,t}}\eta_{i,t,s}=0)$, with differing and
unknown precision parameters $\delta_i$ and $\tau_{i}$ for each
individual $i$. Here $\mathbf{I}$ is an $S_{i,t} \times S_{i,t}$
identity matrix. The multivariate normal distribution for
${\eta}_{i,t}$ is actually a singular multivariate normal
distribution because it is conditioned on the sum of the day's test
effects being zero, done to remove any possibility of confounding
with the daily random effects. (In analysis and computation, this
singular multivariate normal distribution is replaced by the
corresponding lower-dimensional nonsingular multivariate normal
distribution.)

Finally, at the observation level, the dichotomous test data is
modeled as
\begin{eqnarray*}
&&\operatorname{ Pr}(X_{i,t,s,l}=1 \mid\theta_{i,t},a_{i,t,s},
\varphi_{i,t},\eta_{i,t,s},\varepsilon_{i,t,s,l})
\\
&&\qquad= \mathrm{F}(\theta_{i,t}-d_{i,t,s,l}+\varphi_{i,t}+
\eta_{i,t,s})
\\
&&\qquad=\mathrm{ F}(\theta_{i,t}-a_{i,t,s}+\varphi_{i,t}+
\eta_{i,t,s}+\varepsilon_{i,t,s,l}),
\end{eqnarray*}
where $\theta_{i,t}$ represents the $i$th person's ability on day
$t$; we are thus assuming that a person's ability is constant over a
given day, although there could be random fluctuations captured by
the $\varphi_{i,t}$ and $\eta_{i,t,s}$. Letting $\mathrm{ F(\cdot)}$ be
the logistic c.d.f., as previously discussed, results in
%
\begin{eqnarray}\label{irt-12}
&&\operatorname{ Pr}(X_{i,t,s,l}=1 \mid\theta_{i,t},a_{i,t,s},
\varphi_{i,t},\eta_{i,t,s},\varepsilon_{i,t,s,l})
\nonumber
\\[-8pt]
\\[-8pt]
\nonumber
&&\qquad=\frac{\exp(\theta_{i,t}-a_{i,t,s}+\varphi_{i,t}+\eta_{i,t,s}+\varepsilon
_{i,t,s,l})} {
1+\exp(\theta_{i,t}-a_{i,t,s}+\varphi_{i,t}+\eta_{i,t,s}+\varepsilon
_{i,t,s,l})}  .
\end{eqnarray}

\subsection{The system equation in DIR models}\label{sec2.2}
As mentioned in the \hyperref[sec1]{Introduc-} \hyperref[sec1]{tion}, both parametric growth models
and Markov chain models have been utilized in contexts similar to
that of this paper. Here we combine these
ideas, through a generalization of dynamic linear models, to model
an individual's ability growth trajectory over time. The proposed model is
%
\begin{equation}\label{irt-13}
\theta_{i,t}=\theta_{i,t-1}+c_i(1-\rho
\theta_{i,t-1})\Delta_{i,t}^{+}+w_{i,t},
\end{equation}
which has three terms, modeling how current ability, $\theta_{i,t}$
for the $i$th person on the $t$th day, relates to past ability and
other factors. The first term is simply ability at the previous time
point, $\theta_{i,t-1}$.

The second term is a parametric growth model. Here $c_i$ can be
thought of as the average growth rate of the $i$th person's ability
over time and $\Delta_{i, t}^{+}$ is the time lapse between the
person's $t$th test day and $(t-1)$th test day but truncated by a
pre-specified maximum time interval $\Delta_{T_{\max}}$, that is,
$\Delta_{i, t}^{+}=\min\{\Delta_{i, t}, \Delta_{T_{\max}}\}$; thus,
$c_i\Delta_{i,t}^{+}$ would reflect the ability growth over the
given time interval if the growth was indeed linear. However, this
growth is truncated at $\Delta_{T_{\max}}$ (chosen herein to be 14
days), reflecting the fact that, when on vacation, the student's
ability may not be growing. Furthermore, the growth rate often
declines as ability increases (indeed ability typically eventually
plateaus), so that a linear growth model is often unsuitable when
$\theta_{i,t}$ becomes large. The ``correction factor,''
$-\rho\theta_{i,{t-1}}$ in (\ref{irt-13}), compensates for this
effect, slowing down the linear growth as the ability level becomes
larger. $\rho$ is the parameter controlling the rate of this
adjustment, and could be known or unknown. In our testbed example,
$\rho$ is known, based on experiments conducted at MetaMetrics
[Hanlon et al. (\citeyear{Hanlon})]. In principle, $\rho$ should be
individual-specific, but it is distinguishable from $c_i$ only as
the individual's ability level is reaching maturation; our
investigation of ability growth in the testbed data focuses on early
age students, so only the $c_i$ are made individual-specific.

As in all dynamic linear models, the third term, $w_{i,t}$ in
(\ref{irt-13}), represents the random component of the change in
the $i$th person's ability on the $t$th day. We assume it is
$\mathcal{N}(0,\phi^{-1}\Delta_{i, t})$, where $\phi$ is unknown.
Note that this presumes that the random component of a person's
ability change has the variance proportional to the time period
between test days. Note, also, that we suppose that $\phi$ is common
across individuals. The reason for this is clear from
(\ref{irt-12}), in which $\varphi_{i,t}\sim
\mathcal{N}(0,\delta_i^{-1})$ have individual-specific $\delta_i$;
there would be a substantial risk of confounding in the likelihood
between $\delta_i$'s and $\phi^{-1}\Delta_{i,t}$ if the time lapse
between tests for the student were equally spaced.

It is possible to rewrite (\ref{irt-13}) as a first-order Markov
process, and this is beneficial for computational reasons. Indeed,
letting $\lambda_{i,t}=\theta_{i,t}-\rho^{-1}$ and
$g_{i,t}=1-c_i\rho\Delta_{i, t}^{+}$, the system equation
(\ref{irt-13}) becomes
%
\begin{equation}\label{irt-14}
\lambda_{i,t}=g_{i,t}\lambda_{i,t-1}+w_{i,t} ,
\end{equation}
where $w_{i,t} \sim\mathcal{N}(0,\phi^{-1}\Delta_{i,t})$, and this
is in the form of a standard dynamic linear model. (Note that $c_i$
and $\phi$ need to be known for this reduction.)

\subsection{DIR model summary}\label{sec2.3}
To sum up, the one-parameter DIR model is constructed in two
levels as follows:
\begin{eqnarray*}
\mbox{System equation:}&&\qquad \theta_{i,t}=\theta_{i,t-1}+c_i(1-
\rho\theta_{i,t-1})\Delta_{i, t}^++w_{i,t} ,
\\
\mbox{Observation equation:}&&\qquad \operatorname{ Pr}(X_{i,t,s,l}=1 \mid
\theta_{i,t},a_{i,t,s},\varphi_{i,t},\eta_{i,t,s},
\varepsilon_{i,t,s,l})
\\
&&\qquad\qquad=\frac{\exp(\theta_{i,t}-a_{i,t,s}+\varphi_{i,t}+\eta_{i,t,s}+\varepsilon
_{i,t,s,l})} {
1+\exp(\theta_{i,t}-a_{i,t,s}+\varphi_{i,t}+\eta_{i,t,s}+\varepsilon
_{i,t,s,l})} ,
\end{eqnarray*}
where $w_{i,t}\sim\mathcal{N}(0,\phi^{-1}\Delta_{i, t})$,
$\varepsilon_{i,t,s,l} \sim\mathcal{N}(0,\sigma^2)$,
$\varphi_{i,t}\sim\mathcal{N}(0,\delta_i^{-1})$, ${\eta}_{i,t} \sim
\mathcal{N}_{S_{i,t}}(0,\break\tau_i^{-1}\mathbf{I} \mid\sum
_{s=1}^{S_{i,t}}\eta_{i,t,s}=0)$, and $\Delta_{i,
t}^{+}=\min\{\Delta_{i, t}, \Delta_{T_{\max}}\}$, with the $a_{i,t,s}$,
$\rho$,
$\Delta_{i,t}$, $\Delta_{T_{\max}}$ and $\sigma$ being known and
$\theta_{i,t}$, $c_i$, $\phi$, $\delta_i$ and $\tau_i$ being
unknown.

\section{Statistical inference for DIR models}\label{sec3}
In this section the Bayesian methods that will be used for
statistical inference in DIR models are described. Computation is
based on a Gibbs sampling scheme, in conjunction with forward
filtering and backward sampling.

\subsection{Prior distributions for the unknown parameters}\label{sec3.1}
Prior distributions in a Bayesian analysis must be specified
carefully, but they can be either evidence-based priors, reflecting
scientific knowledge of the system under study, or they can be
objective priors, reflecting a lack of such knowledge but possessing
good overall properties---for example, good frequentist properties [see,
e.g., \citet{Berger}]; a~mix of both will be used in the analysis
herein. Specification of evidence-based priors is, of course,
context dependent and, here, will be done within the context of the
MetaMetrics testbed application.

A natural choice of the prior distribution for an individual's
initial latent ability, $\theta_{i,0}$, is
\[
\theta_{i,0} \sim\mathcal{N}(\mu_{G_{j_i}},V_{G_{j_i}}) ,
\]
where $\mu_{G_{j_i}}$ and $V_{G_{j_i}}$ are the mean and the
variance, on a logit scale, of the population ($j$) to which the
individual $i$ belongs---for instance,
the individual's grade in school for the testbed application. For the average
growth rate $c_i$ in system equation (\ref{irt-13}), the natural
objective prior is a constant prior (since $c_i$ is a linear
parameter), but we constrain $c_i$ to be positive, reflecting the
belief that there is a positive learning rate; thus, we choose the
prior
\[
\pi(c_i)\propto I(c_i>0) \qquad\mbox{for all $i$}.
\]
Although $\phi$ is a scale parameter, it occurs at the system-level
of the two-stage model and, hence, the usual scale objective prior
($1/\phi$) would result in an improper posterior; the
computationally simplest adjustment is to use
$\pi(\phi)=1/\phi^{3/2}$, which does result in a proper posterior.
Similarly, for the scale parameters $\delta_i$ and $\tau_i$ we
utilize the objective priors $\pi(\delta_i) = 1/\delta_{i}^{3/2}$
and $\pi(\tau_i) = 1/\tau_{i}^{3/2}$. A natural alternative would be
to try to ``borrow information'' across individuals, by utilizing
gamma hyperpriors for the $\delta_i$'s and $\tau_i$'s.
This complicates the computation, however, and does not
seem necessary for the testbed application.

\subsection{Posterior distribution}\label{sec3.2}
To facilitate the use of Gibbs sampling techniques in computation,
we utilize a mixture of normals representation of the logistic
distribution. From Andrews and Mallow (\citeyear{Andrews}), if $Y$ has a logistic
distribution with location parameter 0 and scale $\pi^2/3$
($\mathcal{L}(0,\frac{\pi^2}{3}))$, one can write the density as
%
\begin{equation}\label{log-density}
f(y)=\frac{e^{-y}}{(1+e^{-y})^2}=\int_0^{\infty} \biggl[
\frac{1}{\sqrt{2\pi}} \frac{1}{2\nu}\exp\biggl\{-\frac{1}{2}\biggl(
\frac{y}{2\nu}\biggr)^2 \biggr\} \biggr]\pi(\nu) \,d\nu,
\end{equation}
where $\nu$ has the Kolmogorov--Smirnov (K--S) density
%
\begin{equation}\label{ks-11}
\pi(\nu)=8\sum_{\alpha=1}^{\infty}(-1)^{(\alpha+1)}
\alpha^2\nu\exp\bigl\{-2\alpha^2\nu^2\bigr\},\qquad
\nu\geq0 .
\end{equation}
Note that the density in square brackets in (\ref{log-density}) is
$\mathcal{N}(0, 4\nu^2)$. By using the idea of data augmentation
from \citet{Tanner},\vadjust{\goodbreak} we consider the latent variable
$Y_{i,t,s,l}$ for each response variable $X_{i,t,s,l}$, where
$Y_{i,t,s,l}
\sim\mathcal{N}(\theta_{i,t}-a_{i,t,s}+\varphi_{i,t}+\eta
_{i,t,s}+\varepsilon_{i,t,s,l},
4\nu_{i,t,s,l}^2)$ and define $X_{i,t,s,l}=1$ if $Y_{i,t,s,l}>0$ and
$X_{i,t,s,l}=0$ otherwise. It is then easy to show that $\operatorname{
Pr}(X_{i,t,s,l}=1|\theta_{i,t}, a_{i,t,s},
\varphi_{i,t},\eta_{i,t,s},\varepsilon_{i,t,s,l})=\exp(\theta_{i,t}-a_{i,t,s}+
\varphi_{i,t}+\eta_{i,t,s}+\varepsilon_{i,t,s,l})/
(1+\break\exp(\theta_{i,t}-a_{i,t,s}+\varphi_{i,t}+\eta_{i,t,s}+\varepsilon_{i,t,s,l}))$,
so that the introduction of the latent variables $Y_{i,t,s,l}$ will
not alter the model (except that there are now formally many more
unknown parameters).

As $\varepsilon_{i,t,s,l}\stackrel{\mathrm{i.i.d.}} \sim
\mathcal{N}(0,\sigma^2)$, it can be marginalized out in the
distribution of $Y_{i,t,s,l}$, resulting in $Y_{i,t,s,l}
\sim\mathcal{N}(\theta_{i,t}-a_{i,t,s}+\varphi_{i,t}+\eta_{i,t,s},4\nu
_{i,t,s,l}^2+\sigma^2)$.
Therefore, the one-parameter DIR models (\ref{irt-12}) and
(\ref{irt-13}) can be rewritten, with latent variables
$\{Y_{i,t,s,l}\}$, as
%
\begin{eqnarray}
\theta_{i,t}&=&\theta_{i,t-1}+c_i(1-\rho
\theta_{i,t-1})\Delta_{i,t}^{+}+w_{i,t},
\label{sm-11}
\\[-2pt]
Y_{i,t,s,l} &=&\theta_{i,t}-a_{i,t,s}+\varphi_{i,t}+
\eta_{i,t,s}+\xi_{i,t,s,l}, \label{sm-12}
\\[-2pt]
\nu_{i,t,s,l}&\sim& \mbox{K--S distribution}, \label{sm-13}
\end{eqnarray}
where $w_{i,t}\sim\mathcal{N}(0,\phi^{-1}\Delta_{i, t})$,
$\varphi_{i,t}\sim\mathcal{N}(0,\delta_i^{-1})$, ${\eta}_{i,t} \sim
\mathcal{N}_{S_{i,t}}(0,\tau_i^{-1}\mathbf{I} \mid
\sum_{s=1}^{S_{i,t}}\eta_{i,t,s}=0)$, and $\xi_{i,t,s,l}\sim
\mathcal{N}(0,\psi_{i,t,s,l}^{-1})$ with
$\psi_{i,t,s,l}^{-1}=4\nu_{i,t,s,l}^2+\sigma^2$.

Define $\theta=(\theta_1,\ldots, \theta_n)'$, where
$\theta_i=(\theta_{i,0},\theta_{i,1},\ldots,\theta_{i,T_i})'$ for
$i=1,\ldots,n$; $c=(c_1,\ldots, c_n)'$ and
$\tau=(\tau_1,\ldots,\tau_n)'$;
$Y=\{Y_{i,t,s,l}\}$, $\nu=\{\nu_{i,t,s,l}\}$ and $X=\{X_{i,t,s,l}\}$
for $l=1,\ldots,K_{i,t,s}$, $s=1,\ldots, S_{i,t}$, $t=1,\ldots, T_i$
and $i=1,\ldots, n$; $\varphi=\{\varphi_{i,t}\}$ for $t=1,\ldots,
T_i$, $i=1,\ldots,n$; $\eta=\{\eta_{i,t,s}\}$ for $s=1,\ldots,
S_{i,t}$, $t=1,\ldots, T_i$ and $i=1,\ldots, n$ and
$\eta_{i,t}^*=(\eta_{i,t,1},\ldots,\eta_{i,t, S_{i,t}-1})'$. Then
the joint posterior density of $\theta$, $Y$, $c$, $\tau$,
$\varphi$, $\eta$, $\nu$ and $\phi$ given the data $X$, in the
one-parameter DIR model, is proportional to
%
\begin{eqnarray}\label{sm-14}
&&\hspace*{-4pt} \pi(\theta,Y,c,\tau,\varphi, \eta, \nu, \phi\mid X)
\nonumber\\[-2pt]
&&\hspace*{-4pt}\qquad\propto \Biggl\{\prod_{i=1}^n\pi(
\theta_{i,0})\pi(c_i)\pi(\delta_i)\pi(
\tau_i) \Biggr\}\pi(\phi) \Biggl\{\prod_{i=1}^n
\prod_{t=1}^{T_i} \prod
_{s=1}^{S_{i,t}}\prod_{l=1}^{K_{i,t,s}}
\pi(\nu_{i,t,s,l}) \Biggr\}
\nonumber
\\[-2pt]
&&\hspace*{-4pt}\qquad\quad{}\times \Biggl\{\prod_{i=1}^n \prod
_{t=1}^{T_i} \prod
_{s=1}^{S_{i,t}}\prod_{l=1}^{K_{i,t,s}}
\bigl(I\{Y_{i,t,s,l}>0\}I\{X_{i,t,s,l}=1\}\nonumber\\
&&\hspace*{-4pt}\hspace*{87pt}{}\qquad\quad+I\{Y_{i,t,s,l}\leq0\}I
\{X_{i,t,s,l}=0\} \bigr)
\nonumber\\[-2pt]
&&\hspace*{-4pt}\hspace*{81pt}\qquad\quad{}\times  \sqrt{\frac{\psi_{i,t,s,l}}{2\pi}}\nonumber
\\[-9pt]
\\[-9pt]
\nonumber
&&\hspace*{-4pt}\hspace*{81pt}\quad\qquad{}\times\exp\biggl(-\frac{\psi
_{i,t,s,l}(Y_{i,t,s,l}-
\theta_{i,t}+a_{i,t,s}-\varphi_{i,t}-\eta_{i,t,s})^2}{2}\biggr)\\[-2pt]
&&\hspace*{-4pt}\hspace*{190pt}\qquad\quad{}\times I\Biggl\{
\eta_{i,t,S_{i,t}}=-\sum_{s=1}^{S_{i,t}-1}
\eta_{i,t,s}\Biggr\} \Biggr\}
\nonumber\\
&&\hspace*{-4pt}\qquad\quad{}\times  \Biggl\{\prod_{i=1}^n \prod
_{t=1}^{T_i}\biggl(\frac{\tau_i}{2
\pi}
\biggr)^{{(S_{i,t}-1)}/{2}}\exp\biggl(-\frac{\tau_i{\eta
_{i,t}^{*\prime}}\Sigma_{i,t}^{-1}\eta_{i,t}^*}{2}\biggr) \Biggr\} \nonumber\\
&&\hspace*{-4pt}\qquad\quad{}\times\Biggl\{
\prod
_{i-1}^n \prod
_{t=1}^{T_i}\sqrt{\frac{\delta_i}{2\pi}}\exp\biggl(-
\frac{\delta_i\varphi_{i,t}^2}{2}\biggr) \Biggr\}
\nonumber
\\
&&\hspace*{-4pt}\qquad\quad{}\times  \Biggl\{\prod_{i=1}^n \prod
_{t=1}^{T_i} \sqrt\frac{\phi}{2\pi\Delta_{i,
t}}\exp
\biggl(-\frac{\phi\{\theta_{i,t}-\theta_{i,t-1}-c_i(1-\rho\theta
_{i,t-1})\Delta_{i,t}^{+}\}^2}{2\Delta_{i,
t}}\biggr) \Biggr\},
\nonumber
\end{eqnarray}
where
\[
\Sigma_{i,t}^{-1}=
\pmatrix{
2& 1 &\cdots&1
\vspace*{2pt}\cr
1 & 2& \cdots& 1
\vspace*{2pt}\cr
\vdots& \vdots& \ddots& \vdots
\vspace*{2pt}\cr
1 & 1 & \cdots& 2}_{(S_{i,t}-1)\times(S_{i,t}-1)},
\]
and $I(Z\in A)$ is the indicator function equal to $1$ if the random
variable $Z$ is contained in the set $A$; $\pi(\theta_{i,0})$,
$\pi(c_i)$, $\pi(\delta_i)$, $\pi(\tau_i)$, $\pi(\phi)$ are the
priors specified in the previous subsection, and
$\pi(\nu_{i,t,s,l})$ is the K--S density defined at the beginning of
this subsection. This is a proper posterior under very mild conditions;
see Appendix~\ref{appC}.

\subsection{Computation}\label{sec3.3}

Computation is done by a MCMC scheme that samples from the posterior
(\ref{sm-14}) via a block Gibbs sampling scheme, utilizing the
forward filtering and backward sampling algorithm at a key point.
The steps of the algorithm are given in Appendix~\ref{appA}.

From the MCMC samples, statistical inferences are straightforward.
For example, an estimate and $95\%$ credible interval for the latent
ability trait $\theta_{i,t}$ can be formed from the median, $2.5\%$,
and $97.5\%$ empirical quantiles of the corresponding MCMC
realizations. In examples, these will be graphed as a function of
$t$ so that the adaptive nature of the model is apparent.

\section{Simulated examples}\label{sec4}
In this section a simulated example is used to illustrate the
inferences from the proposed one-parameter DIR models and to study
their properties, primarily from a frequentist perspective.

The simulation examines the model's behavior for multiple
individuals taking a series of tests that are scheduled during
different time periods. In particular, suppose there are 10
individuals and each individual has taken tests on 50 different
days. Thus, $n=10$ and $T_i=50$, for $i=1,\ldots,10$. During each
distinctive test day, the individual takes four tests; thus,
$S_{i,t}=4$ for $t=1,\ldots,50$, $i=1,\ldots,10$. Each test consists
of 10 items, so that $K_{i,t,s}=10$ for $s=1,\ldots, 4$,
$t=1,\ldots, 50$ and $i=1,\ldots, 10$. For the $i$th person, the
time lapse between two different\vadjust{\goodbreak} tests is assumed to be a function
of the $t$th day, that is, $\Delta_{i, t}=10+t$, for $i=1,\ldots,10$,
$t=1,\ldots,T_i/2$ and $\Delta_{i, t}=t-10$, for
$t=T_i/2,\ldots,T_i$. Finally, the unknown values of parameters in
the models are chosen as follows:
\begin{itemize}
\item$\phi=1/0.0218^2$, and the corresponding standard deviation of
the random component $w_{i,t}$ in
system equation (\ref{irt-13}) is $0.0218\sqrt{\Delta_{i,t}}$.
\item
$c=(0.0055,0.0065,0.0026,0.0037,0.0061,0.0047,0.0035,0.0043,0.0039,\break 0.0015)'$,
where each element in the vector $c$ corresponds to the
$i$th person's average growth rate, respectively, for
$i=1,\ldots,10$.
\item$\delta=(2.0408,1.3333,1.8182,1.2346,1.5873,1, 2.2222,1.0526, 1.1494,
2)'$,\break where each element in the vector $\delta$ corresponds to
the precision parameter of daily random effects for the $i$th
person, respectively, $i=1,\ldots,10$.
\item$\tau=(4, 3.1250, 4.3478, 2.7027, 3.7037, 2.8571, 4, 2.2222,
9.0909, 4.5455)'$,\break
where each element in the vector $\tau$ corresponds to the
precision parameter of test random effects for the $i$th
person, respectively, $i=1,\ldots,10$.
\end{itemize}
According to the observation equation (\ref{irt-12}), we then
simulated values for the unknown variables and set the test
difficulties, $a_{i,t,s}$, to be $\theta_{i,t}+\zeta$, where $\zeta$
is a random variable with uniform distribution on ($-0.1,0.1$). The
values of $\varepsilon_{i,t,s,l}$ were drawn from
$\mathcal{N}(0,0.7333^2)$ and the value of $0.7333$ is used in the
test design for MetaMetrics. Finally, we chose $\rho=0.1180$, which
is the value estimated by MetaMetrics in their studies [\citet{Hanlon}].

\begin{figure}

\includegraphics{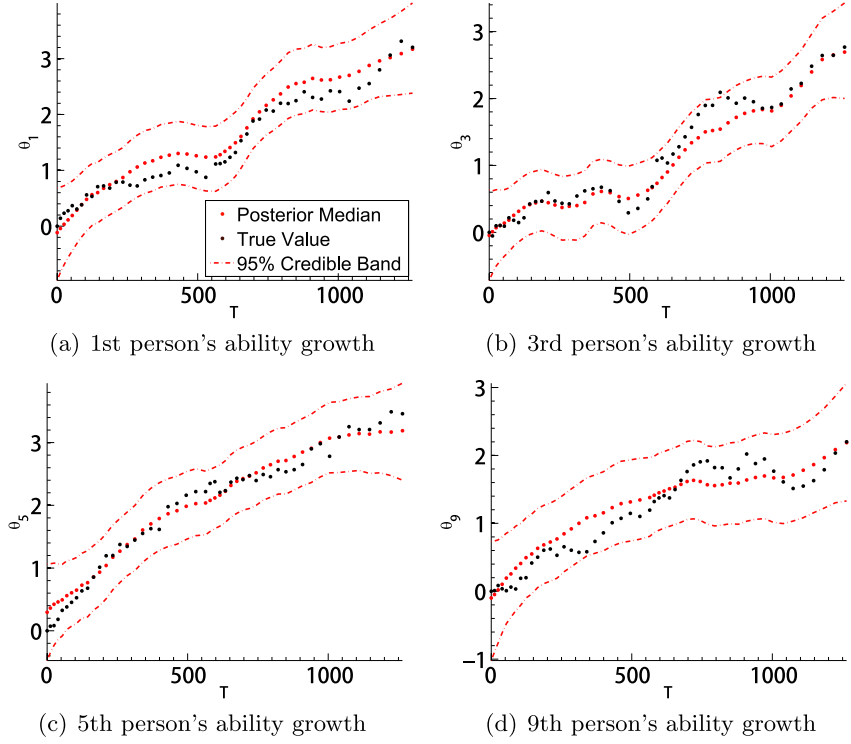}\vspace*{-3pt}

\caption{Estimated and actual ability trajectories of 4
individuals from the simulated data.}\label{fig-22}\vspace*{-3pt}
\end{figure}

From dichotomous data obtained from the simulation, the Bayesian
machinery from Section~\ref{sec3} was used in estimating the model parameters
in (\ref{irt-12}) and (\ref{irt-13}). Figure~\ref{fig-22}
shows estimates of the ability trajectory for the $1$st, $3$rd,
$5$th and $9$th individuals. The red dots in the figures correspond
to the estimated posterior median of the ability $\theta_{i,t}$ at
the $t$th day for the $i$th person, and the red dashed lines give
the 2.5\% and 97.5\% quantile trajectories of $\theta_{i,t}$, for
$t=1,\ldots,50$. The black dots are the real abilities at the $t$th
day for the $i$th person in the simulation. The third trajectory is
typical of what is expected in terms of increasing ability, and is
smoothly handled by the Bayesian machinery. The other three
trajectories are highly nonmonotonic; the Bayesian estimates err in
trying to be increasing (as they are designed to do), but do adapt
to the nonmonotonicity when the evidence becomes strong enough.

One method of evaluating the success of the inferential scheme is to
evaluate the percentage of time that the true ability,
$\theta_{i,t}$, is contained in the $95\%$ credible interval of
estimated ability for each individual. For the ten individuals,
these estimated coverages were $100\%$, $100\%$, $99\%$, $99\%$,
$100\%$, $100\%$, $94\%$, $100\%$, $100\%$ and $91\%$, which
produce an overall estimated coverage of $98.3\%$. Thus, while the
inferential method is Bayesian, it seems to be yielding sets that
have good frequentist coverage.

%

%
\begin{figure}[b]
\vspace*{-3pt}
\includegraphics{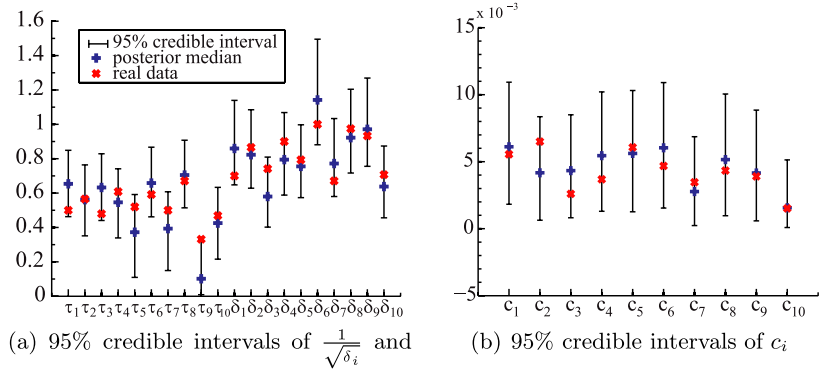}\vspace*{-3pt}

\caption{95\% credible intervals of
$c_i$, $\frac{1}{\sqrt{\tau_i}}$ and $\frac{1}{\sqrt{\delta_i}}$,
for $i=1,\ldots,10$ with the simulated data.} \label{fig-23}
\end{figure}


To summarize the results for the $c_i$'s, $\tau_i^{-1/2}$'s and
$\delta_i^{-1/2}$'s , we compare their true values with the
corresponding estimated values in Figure~\ref{fig-23}. In these
plots, the black bar represents the 95\% credible interval of the
posterior distribution. The blue plus stands for the estimated
posterior median and the red cross is the true value in the
simulation. Moreover, the estimated posterior median of
$\phi^{-1/2}$ is $0.0315$\vadjust{\goodbreak} and its 95\% credible interval is
$[0.0148,0.0484]$.
Note that the true values of the $c_i$'s, $\tau_i^{-1/2}$'s,
$\delta_i^{-1/2}$'s and $\phi$ are all contained in the 95\%
credible intervals except $\tau_9^{-1/2}$; thus, the empirical
coverage for these parameters is $96.77\%$.

%
\begin{figure}

\includegraphics{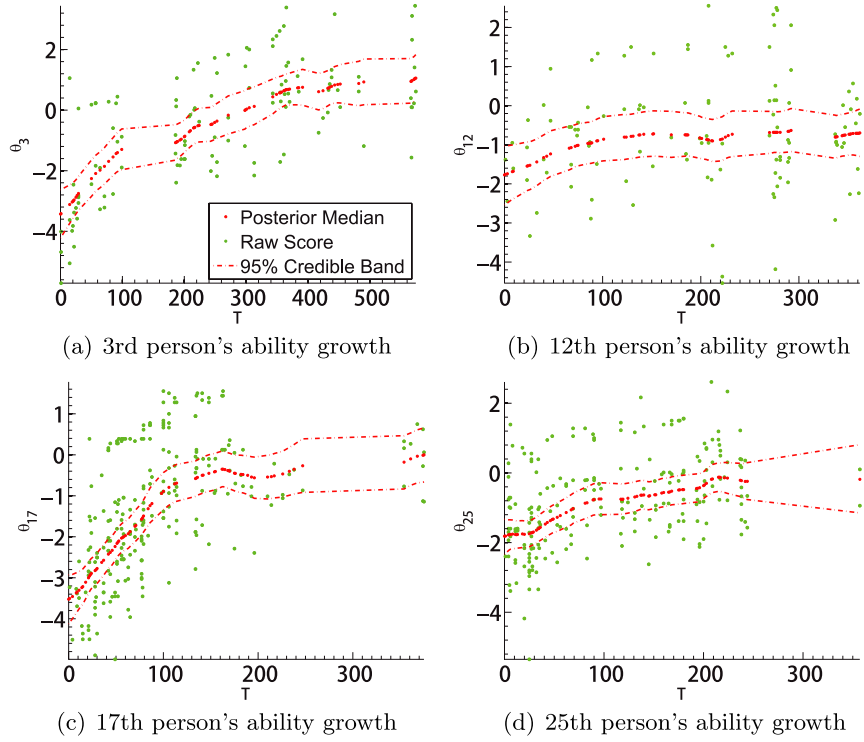}

\caption{Retrospective estimates of ability trajectories of 4
individuals from the
MetaMetrics data.}\label{fig-31}
\end{figure}

%

\section{MetaMetrics testbed}\label{sec5}
In this section we apply the DIR model to the testbed MetaMetrics
data. A sample of $25$ individuals from the data base of students in
certain elementary schools in Mississippi is considered here; the
differing characteristics of the students are described in Appendix
\ref{appB}. The primary focus is study of the goals mentioned in Section~\ref{sec1.2}.

\subsection{Retrospective estimation of ability growth}\label{sec5.1}
First consider retrospective estimation of the reading ability for
an individual, utilizing all the data recorded for that individual.
Figure~\ref{fig-31} presents the resulting growth trajectories for
the $3$rd, $12$th, $17$th and $25$th individuals studied. In Figure
\ref{fig-31} the red dots are the posterior median estimates of
each individual ability and the red dashes correspond to the $2.5\%$
and $97.5\%$ quantiles of the posterior distributions of the
abilities, while the green dots correspond to estimates of an
individual's abilities obtained by solving the equation that the
expectation of expected score for a person's ability is equivalent
to the observed score; these can roughly be thought of as the raw
test scores put on the same scale as the $\theta_{i,t}$. The most
interesting feature of these growth trajectories is that, while
indeed there typically does appear to be overall growth in ability,
this growth need not be monotone. In particular, when there is a
large time gap between subsequent tests, the ability appears to drop
for some individuals. One natural explanation is that, during
vacations, a student may not read and could actually lose ability.
Another possible explanation is that the student has become less
adept at implementation of CAIT after a long break.\looseness=-1

%

%
\begin{figure}

\includegraphics{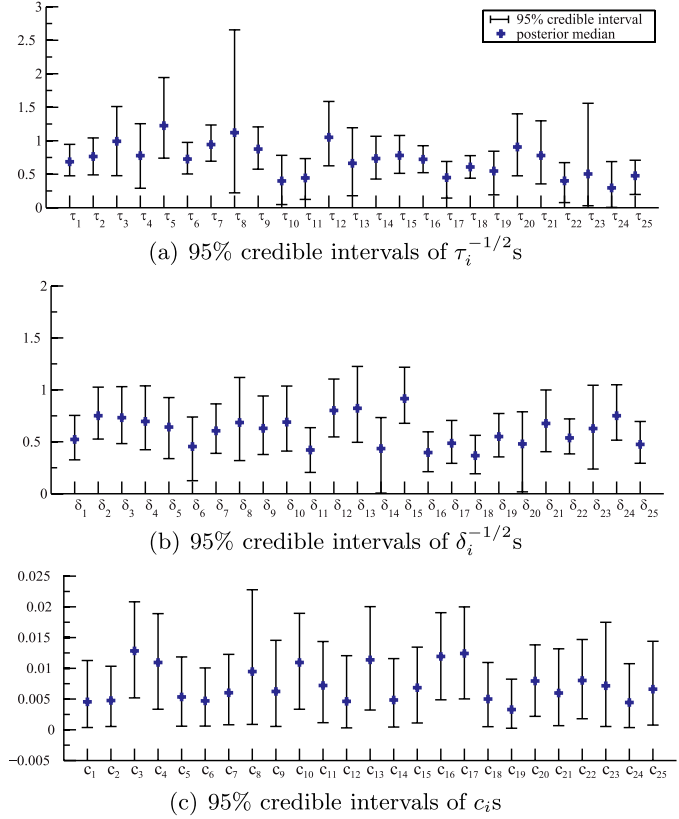}

\caption{95\% credible intervals of the $\tau_i^{-1/2}$s,
$\delta_i^{-1/2}$'s and $c_i$'s with the MetaMetrics data set.} \label{fig-24}
\end{figure}

%

Figure~\ref{fig-24} gives the summaries of the posterior
distributions of the standard deviations of test random effects,
$\tau_i^{-1/2}$'s, the standard deviations of the daily random
effects, $\delta_i^{-1/2}$'s, and the average growth rates, $c_i$'s,
for $i=1,\ldots, 25$. Moreover, the estimated posterior median of
$\phi^{-1/2}$ is $0.0612$ and its 95\% credible interval is
$[0.0477,0.0757]$.

Figures~\ref{fig-24}(a) and (b) show that the standard deviations
of two random effects are almost all quite large with 95\% credible
intervals well separated from zero. Recall that these were included
in the model to account for a possible lack of the local
independence; the evidence is thus strong that the local
independence is, indeed, not tenable for this data and that both
types of random effects are present. The consistency of the standard
deviations of the random effects across individuals is somewhat
surprising, but lends credence to the notion that random effect
modeling of the local dependence is fruitful.

\subsection{On-line estimation of ability growth}\label{sec5.2}
In on-line estimation of reading ability, essentially the same model
is used, but, at each time point, only the data up to that time is
utilized. Instead of having $\phi^{-1/2}$
unknown, however, we utilize $\phi^{-1/2}=0.0612$, the estimate arising from
the retrospective analysis; $\phi^{-1/2}$
cannot be effectively estimated in an on-line mode.

\begin{figure}

\includegraphics{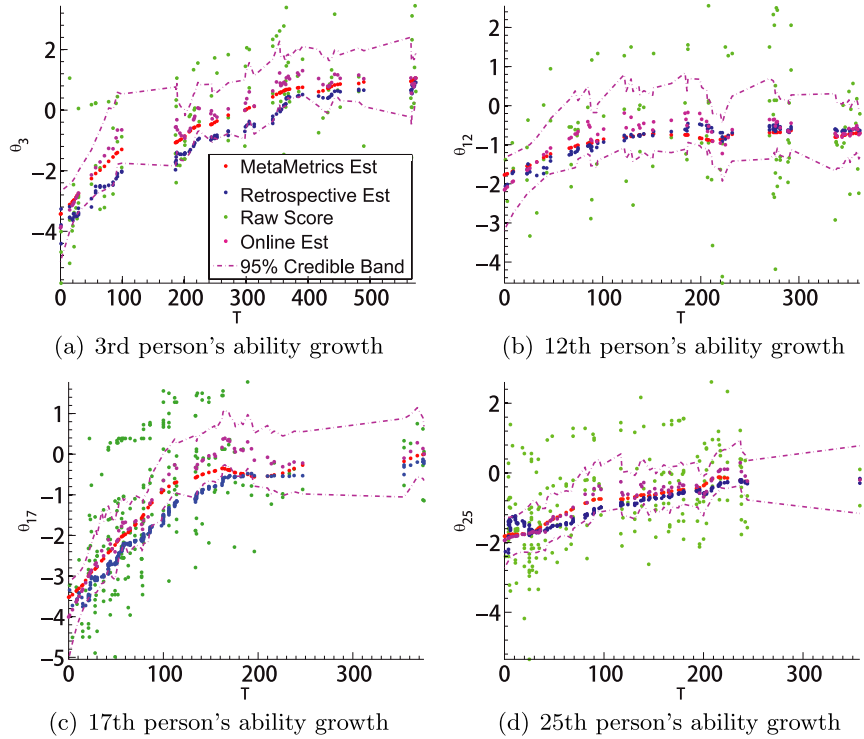}

\caption{On-line estimates of ability trajectories of 4 individuals
from the MetaMetrics data.}\label{fig-41}
\end{figure}

Applying the Bayesian methodology yields on-line posterior median
ability estimates, as well as the $2.5\%$ and $97.5\%$ quantiles of
the posterior distribution of abilities for the $25$ individuals
being studied; these are the purple dots and and dashed purple lines
in Figure~\ref{fig-41}, shown for the $3$rd, $12$th, $17$th and
$25$th individuals. Again, the green dots show the raw score
estimates of each individual ability at each time point, and the red
dots are the retrospective estimates discussed earlier. In these
figures we also include, as blue dots, the ability estimates
obtained from the current methodology of MetaMetrics, which is a
partial Bayesian procedure.

As expected, the on-line ability estimates are much more variable
than the retrospective estimates. Sometimes, the on-line estimates
seem to be somewhat more variable than the current MetaMetrics
estimates (the blue dots). This is because at each online estimation
point, the current methodology of MetaMetrics uses a very tight
prior (arising from the previous data) for the student's ability.

While we do not know the truth here, it is plausible that the
retrospective red dots are our best guesses as to the true
abilities, and we can then judge how well the various on-line
procedures are doing relative to these best guesses. Our on-line
estimates are generally closer to these retrospective estimates than
the current MetaMetrics estimates (the 12th individual being the
interesting exception). In fact, the average mean squared error of
our on-line estimates relative to the retrospective estimates is
$0.0851$, while the average mean squared error of the current
MetaMetrics estimates is $0.1311$.

If we do view the retrospective estimates (red dots) as surrogates
for the truth, it is interesting to see how often these fall outside
the on-line uncertainty bands (purple lines). This happened very
rarely; individual $17$ in Figure~\ref{fig-41} was one case in
which this sometimes happened. One final observation from Figure
\ref{fig-41} is that the current MetaMetric estimates usually are
lower than our on-line estimates of the person's reading ability.

%

%

\section{Conclusions and generalizations}\label{sec6}
The evidence of violation of the local dependence assumption in CAIT
situations is generally strong, and use of test and daily random
effects to model the local dependence seems to be necessary and
successful. Embedding a dynamic linear model framework for an
individual's ability trajectory within the logistic IRT structure
provides a powerful and individually adaptive method for dealing
with longitudinal testing data.

The retrospective DIR model analysis seems excellent for assessing
actual ability trajectories and, hence, is of considerable use in
understanding population behavior, such as the frequently observed
drops in ability after a long pause in testing. The on-line DIR
analysis provides real-time ability estimates for assignments of
material at the right difficulty level and other possible
educational goals.

A key advantage of the Bayesian framework adopted is that
uncertainty in all unknowns can be built into the model (e.g.,
uncertainty in the difficulty of the random test items), and
uncertainty of the estimates is available for all inferences. Also,
prior information (e.g., knowledge about ability distributions over
the population and knowledge that general growth in ability is
expected) can be built into the analysis, in a nondogmatic fashion
that allows the data to overrule the prior.

Many extensions are possible, such as the already mentioned
extension to two-parameter and three-parameter IRT models. If one
also had data for individuals over a period of many years---including years
near the maturation point in one's reading ability---it would be possible to include individual-specific $\rho_i$ in
the model.

\begin{appendix}

\section{The MCMC computation}\label{appA}

The MCMC scheme that will be used to sample from the posterior
(\ref{sm-14}) is a block Gibbs sampling scheme, utilizing the
forward filtering and backward sampling algorithm at a key point.
Because of the block Gibbs sampling scheme, we need only specify the
conditional distributions of a block of variables given the data and
other unknown variables.

\subsection{Sampling Y: Truncated normal distribution
sampling}\label{A-11} Given $\theta$, $\varphi$, $\eta$ and $\nu$,
the latent variables $\{Y_{i,t,s,l}\}$ are sampled from
\begin{eqnarray*}
Y_{i,t,s,l} &\sim&\mathcal{N_+}\bigl(\theta_{i,t}-a_{i,t,s}+
\varphi_{i,t}+\eta_{i,t,s},\psi_{i,t,s,l}^{-1}\bigr)\qquad
\mbox{if } X_{i,t,s,l}=1,
\\
Y_{i,t,s,l} &\sim&\mathcal{N_-}\bigl(\theta_{i,t}-a_{i,t,s}+
\varphi_{i,t}+\eta_{i,t,s},\psi_{i,t,s,l}^{-1}\bigr)\qquad
\mbox{if } X_{i,t,s,l}=0 ,
\end{eqnarray*}
where $\mathcal{N}_{+}$ means the normal distribution truncated at
the left by zero, while $\mathcal{N}_{-}$ is the normal distribution
truncated at the right by zero and
$\psi_{i,t,s,l}^{-1}=4\nu_{i,t,s,l}^2+\sigma^2$. Sampling from
truncated normals is fast and easy.

\subsection{\texorpdfstring{Sampling $\theta$: Forward filtering and backward sampling}
{Sampling theta: Forward filtering and backward sampling}}

The latent ability vector
$\theta_i=(\theta_{i,0},\ldots,\theta_{i,T_i})$, for each
individual, is typically high-dimensional with highly correlated
coordinates, so sampling of the variables would appear to be highly
challenging. To overcome this roadblock, the proposed model is
transformed so that $\theta_i$ could be block sampled---within a
Gibbs sampling step conditional on the other parameters---by the
highly efficient forward filtering and backward sampling algorithm.

To see this, consider $\phi$, $c$, $Y$, $\varphi$, $\eta$ and $\nu$
as given (the Gibbs sampling step). Define
$Z_{i,t,s,l}=Y_{i,t,s,l}+a_{i,t,s}-\varphi_{i,t}-\eta_{i,t,s}-\rho^{-1}$
and utilize the formulation of the model in (\ref{irt-14}). Then,
the (conditional) one-parameter DIR model fits the framework of
dynamic linear models [\citet{West}], that~is,
\begin{eqnarray*}
\mbox{System equation:} &&\qquad\lambda_{i,t}=g_{i,t}
\lambda_{i,t-1}+w_{i,t},
\\
\mbox{Observation equation:} &&\qquad Z_{i,t,s,l}=\lambda_{i,t}+
\xi_{i,t,s,l},
\end{eqnarray*}
where $w_{i,t} \sim\mathcal{N}(0,\phi^{-1}\Delta_{i, t})$,
$\xi_{i,t,s,l}\sim\mathcal{N}(0,\psi_{i,t,s,l}^{-1})$ with
$\psi_{i,t,s,l}^{-1}=4\nu_{i,t,s,l}^2+\sigma^2$. As indicated in
\citet{West}, the forward filtering and backward
sampling algorithm to block update each vector $\theta_i$ proceeds
as follows.

Since $\lambda_{i,0}=\theta_{i,0}-\rho^{-1}$ and $\theta_{i,0}\sim
\mathcal{N}(\mu_{G_j},V_{G_j})$, the conditional prior for
$\lambda_{i,0}$ is $\lambda_{i,0}\sim
\mathcal{N}(\mu_{G_j}-\rho^{-1}, V_{G_j})$. Define information\vadjust{\goodbreak}
available on the $t$th day for the $i$th person as
\[
D_{i,t}=\{g_{i,q},\phi,\psi,\varphi,\eta,c, Z_{i,q,1,1},
\ldots,Z_{i,q,S_{i,q},K_{i,q,S_{i,q}}}\}_{q=1}^t.
\]
We claim that the posterior distribution of $\lambda_{i,t}$ is then
%
\begin{equation} \label{sq-11}
\lambda_{i,t} \mid D_{i,t} \sim\mathcal{N}(
\mu_{i,t}, V_{i,t}) ,
\end{equation}
which can be verified by induction as follows. Assume that, on the
$(t-1)$th day, the
posterior of $\lambda_{i,t-1}$, given $D_{i,t-1}$, is
$\mathcal{N}(\mu_{i,t-1}, V_{i,t-1})$. And it is easy to see this
assumption is true when $t=1$. Then, from the system equation, it is
easy to establish that $\lambda_{i,t} \mid D_{i,t-1} \sim
\mathcal{N}(d_{i,t}, R_{i,t})$ is a prior for $\lambda_{i,t}$, where
$d_{i,t}=g_{i,t}\mu_{i,t-1}$ and $R_{i,t}=g_{i,t}^2
V_{i,t-1}+\phi^{-1}\Delta_{i, t}$. Therefore, we have
\begin{eqnarray*}
\operatorname{ Pr}(\lambda_{i,t} \mid D_{i,t})&\propto& \operatorname{ Pr}(
\lambda_{i,t} \mid D_{i,t-1}) \prod_{s=1}^{S_{i,t}}
\prod_{l=1}^{K_{i,t,s}}\operatorname{ Pr}(Z_{i,t,s,l}
\mid\lambda_{i,t})
\\
&\propto& \exp\biggl\{-\frac{R_{i,t}^{-1}(\lambda
_{i,t}-d_{i,t})^2}{2}\biggr\} \\
&&{}\times\Biggl\{\prod
_{s=1}^{S_{i,t}} \prod_{l=1}^{K_{i,t,s}}
\exp\biggl\{-\frac{\psi_{i,t,s,l}(Z_{i,t,s,l}-\lambda
_{i,t})^2}{2}\biggr\} \Biggr\}.
\end{eqnarray*}
Then, at the $t$th day, the posterior distribution of $\lambda_{i,t}$
is as (\ref{sq-11}), where $\mu_{i,t}=V_{i,t}(R_{i,t}^{-1}d_{i,t}+
\sum_{s=1}^{S_{i,t}}\sum_{l=1}^{K_{i,t,s}}\psi_{i,t,s,l}Z_{i,t,s,l})$
and
$V_{i,t}=(\sum_{s=1}^{S_{i,t}}\sum_{s=1}^{K_{i,t,s}}\psi
_{i,t,s,l}+\break R_{i,t}^{-1})^{-1}$.

The above updating procedure is called forward filtering and after
it is complete and all quantities, that is, $\mu_{i,t}$ and $V_{i,t}$
are saved, we can begin the backward sampling of $\lambda_{i,t}$.
For the time $t=T_i$, we sample $\lambda_{i,t}$ directly from
$\mathcal{N}(\mu_{i,T}, V_{i,T})$. As the time from $t=(T_i-1)$ to
$0$, at each time we draw $\lambda_{i,t}$ from
\[
\lambda_{i,t} \mid\lambda_{i,t+1}, D_{i,t} \sim
\mathcal{N}(h_{i,t}, H_{i,t}),
\]
where $h_{i,t}=H_{i,t}(V_{i,t}^{-1}\mu_{i,t}+\phi
g_{i,t+1}\Delta_{i, t+1}^{-1} \lambda_{i,t+1})$ and $H_{i,t}=(\phi
g_{i,t+1}^2 \Delta_{i,t+1}^{-1}+V_{i,t}^{-1})^{-1}$. This follows
from
\begin{eqnarray*}
\operatorname{ Pr}(\lambda_{i,t} \mid\lambda_{i,t+1}, D_{i,t})&
\propto& \operatorname{ Pr}(\lambda_{i,t} \mid D_{i,t})\operatorname{ Pr}(
\lambda_{i,t+1} \mid\lambda_{i,t}, D_{i,t})
\\
&\propto& \exp\biggl\{-\frac{V_{i,t}^{-1}(\lambda_{i,t}-\mu
_{i,t})^2}{2}\biggr\} \\
&&{}\times \exp\biggl\{-
\frac{\phi\Delta_{i,t+1}^{-1}(\lambda_{i,t+1}-g_{i,t+1}\lambda
_{i,t})^2}{2}\biggr\}.
\end{eqnarray*}
Thus, for $t=0,\ldots,T_i$, we set
$\theta_{i,t}=\lambda_{i,t}+\rho^{-1}$ and each vector $\theta_{i}$
is sampled as a whole block, noticing that
\[
\operatorname{ Pr}(\theta_i \mid D_{i,T_i})=\operatorname{ Pr}(
\theta_{i,T_i} \mid D_{i,T_i})\operatorname{ Pr}(\theta_{i,{T_i-1}} \mid
\theta_{i,T_i},D_{i,T-1})\cdots
\operatorname{ Pr}(
\theta_{i,0} \mid\theta_{i,1},D_{i,0}).
\]

\subsection{Sampling c: Truncated normal distribution sampling}
When $\theta$ and $\phi$ are given, the full conditional
distribution of $c_i$ is the truncated normal distribution
\begin{eqnarray*}
&&c_i\sim\mathcal{N}_{+} \biggl(\frac{\sum_{t=1}^{T_i}(1-\rho\theta
_{i,t-1})(\theta_{i,t}-\theta_{i,t-1})\Delta_{i,t}^+
\Delta_{i,t}^{-1}}{\sum_{t=1}^{T_i}
(\Delta_{i,t}^+(1-\rho\theta_{i,t-1}))^2\Delta_{i,t}^{-1}},\\
&&\hspace*{84pt}{} \frac{1}{\phi\sum_{t=1}^{T_i}(\Delta_{i,t}^+
(1-\rho\theta_{i,t-1}))^2\Delta_{i,t}^{-1}} \biggr).
\end{eqnarray*}

\subsection{\texorpdfstring{Sampling $\eta$: Multivariate normal distribution sampling}
{Sampling eta: Multivariate normal distribution sampling}}
When $\theta$, $\varphi$, $\tau$, $Y$ and $\nu$ are given, if
$S_{i,t}>1$, then the full conditional distribution of
$\eta_{i,t}^*$ is the multivariate normal distribution
\begin{eqnarray*}
\eta_{i,t}^*\hspace*{-0.5pt} \sim\hspace*{-0.5pt}\mathcal{N}_{S_{i,t}-1} \bigl(
\bigl(A_{i,t}^T \Sigma_{\psi_{i,t}}^{-1}
A_{i,t}+\tau_i\Sigma_{i,t}^{-1}
\bigr)^{-1}\hspace*{-0.5pt}A_{i,t}^T \Sigma_{\psi_{i,t}}^{-1}
Y_{i,t}^{*}, \bigl(A_{i,t}^T
\Sigma_{\psi_{i,t}}^{-1} A_{i,t}+\tau_i
\Sigma_{i,t}^{-1}\bigr)^{-1} \bigr) ,
\end{eqnarray*}
where $Y_{i,t}^*=(Y_{i,t,1,1}-\theta_{i,t}+a_{i,t,1}-\varphi_{i,t},
\ldots,
Y_{i,t,1,K_{i,t,1}}-\theta_{i,t}+a_{i,t,K_{i,t,1}}-\varphi_{i,t},
\ldots,\break
Y_{i,t,S_{i,t},K_{i,t,S_{i,t}}}-\theta
_{i,t}+a_{i,t,K_{i,t,S_{i,t}}}-\varphi_{i,t})'$,
$\Sigma_{\psi_{i,t}}^{-1}=\operatorname{
diag}((\psi_{i,t,1,1},\ldots,\break
\psi_{i,t,S_{i,t}, K_{i,t,S_{i,t}}})')$,
\begin{eqnarray*}
A_{i,t}=\pmatrix{
\mathbf{1}_{K_{i,t,1}}&
\mathbf{0} & \cdots&\mathbf{0}
\vspace*{2pt}\cr
\mathbf{0} & \mathbf{1}_{K_{i,t,2}} & \cdots&\mathbf{0}
\vspace*{2pt}\cr
\vdots& \vdots& \ddots& \vdots
\vspace*{2pt}\cr
\mathbf{0} & \mathbf{0} & \cdots&\mathbf{1}_{K_{i,t,S_{i,t}-1}}
\vspace*{2pt}\cr
-\mathbf{1}_{K_{i,t,S_{i,t}}} & -\mathbf{1}_{K_{i,t,S_{i,t}}} & \cdots&-
\mathbf{1}_{K_{i,t,S_{i,t}}}}_{(\sum_{s=1}^{S_{i,t}}K_{i,t,s})\times(S_{i,t}-1)},
\end{eqnarray*}
where $\mathbf{1}_K$ is a $K$-dimensional column vector with each
element being 1 and
$\eta_{i,t,S_{i,t}}=-\sum_{s=1}^{S_{i,t}-1}\eta_{i,t,s}$. When
$S_{i,t}=1$, $\eta_{i,t,S_{i,t}}=0$.

\subsection{\texorpdfstring{Sampling $\tau$: Gamma distribution sampling}
{Sampling tau: Gamma distribution sampling}}
When $\eta$ is given, the full conditional distribution of
$\tau_{i}$ is the gamma distribution
\[
\tau_i \sim\mathcal{G}a \biggl(\frac{\sum_{t=1}^{T_i}S_{i,t}-(T_i+1)}{2},
\frac{\sum_{t=1}^{T_i}{\eta_{i,t}^{*\prime}}\Sigma_{i,t}^{-1}\eta_{i,t}^*}{2}
\biggr) .
\]

\subsection{\texorpdfstring{Sampling $\varphi$: Normal distribution sampling}
{Sampling phi: Normal distribution sampling}}
When $\theta$, $\eta$, $\delta$, $Y$ and $\nu$ are given, the full
conditional distribution of $\varphi_{i,t}$ is the normal
distribution
\begin{eqnarray*}
&&\varphi_{i,t} \sim\mathcal{N} \biggl(\frac{\sum_{s=1}^{S_{i,t}}\sum
_{l=1}^{K_{i,t,s}}\psi_{i,t,s,l}(Y_{i,t,s,l}-\theta_{i,t}+
a_{i,t,s}-\eta_{i,t,s})}{\sum_{s=1}^{S_{i,t}}\sum_{l=1}^{K_{i,t,s}}\psi
_{i,t,s,l}+\delta_i},\\
&&\hspace*{158pt}{}\frac{1}{\sum_{s=1}^{S_{i,t}}\sum_{l=1}^{K_{i,t,s}}\psi_{i,t,s,l}+\delta
_i} \biggr).
\end{eqnarray*}

\subsection{\texorpdfstring{Sampling $\delta$: Gamma distribution sampling}
{Sampling delta: Gamma distribution sampling}}
When $\varphi$ is given, the full conditional distribution of
$\delta_{i}$ is the gamma distribution
\[
\delta_i \sim\mathcal{G}a \biggl(\frac{T_i-1}{2},
\frac{\sum_{t=1}^{T_i}\varphi_{i,t}^2}{2} \biggr) .
\]

\subsection{\texorpdfstring{Sampling $\phi$: Gamma distribution sampling}
{Sampling phi: Gamma distribution sampling}}
When $\theta, c$ is given, the full conditional distribution of
$\phi$ is the gamma distribution
\begin{eqnarray*}
\phi\sim\mathcal{G}a \biggl(\frac{\sum_{i=1}^n T_i-1}{2},\frac{\sum
_{i=1}^n\sum_{t=1}^{T_i}
\Delta_{i,t}^{-1}(\theta_{i,t}-\theta_{i,t-1}-c_i(1-\rho\theta
_{i,t-1})\Delta_{i,t}^{+})^2}{2} \biggr) .
\end{eqnarray*}

\subsection{\texorpdfstring{Sampling $\nu$: Metropolis--Hastings sampling}
{Sampling nu: Metropolis--Hastings sampling}}\label{A-12}

Given $Y$, $\theta$, $\varphi$ and $\eta$, the full conditional
distribution of $\nu_{i,t,s,l}$ is proportional to
\begin{eqnarray*}
\pi(\nu_{i,t,s,l}| Y, \theta, \varphi, \eta)&\propto&\sqrt{
\frac{1}{\sigma^2+4\nu_{i,t,s,l}^2}} \\
&&{}\times \exp\biggl\{-\frac
{(Y_{i,t,s,l}-\theta_{i,t}+a_{i,t,s}-\varphi_{i,t}-\eta_{i,t,s})^2} {
2(\sigma^2+4\nu_{i,t,s,l}^2)} \biggr\},
\end{eqnarray*}
which is not in closed form. So we shall resort to a
Metropolis--Hastings scheme to sample this distribution. A suitable
proposal for sample $\nu$ is the K--S distribution itself. Thus, we first
sample $\nu$ from the K--S distribution whose density is defined in
(\ref{ks-11}). Then, we let
\begin{eqnarray*}
\nu^{(M)}_{i,t,s,l}=\cases{
\nu^{*},
&\quad $\mbox{with probability $\operatorname{ min}(1,LR)$,}$
\vspace*{2pt}\cr
\nu^{(M-1)}_{i,t,s,l}, &\quad $\mbox{otherwise},$}
\end{eqnarray*}
where, given $Y$, $\theta$, $\varphi$ and $\eta$,
\begin{eqnarray*}
LR&=&\sqrt{\frac{\sigma^2+4(\nu^{(M-1)}_{i,t,s,l})^2}{\sigma^2+4(\nu
^*)^2}}\exp\biggl\{-\frac{(Y_{i,t,s,l}-\theta_{i,t}+a_{i,t,s}-\varphi
_{i,t}-\eta_{i,t,s})^2}{2}
\\
&&\hspace*{110pt}{}\times \biggl(\frac{1}{\sigma^2+4(\nu^*)^2}-\frac{1} {
\sigma^2+4(\nu^{(M-1)}_{i,t,s,l})^2} \biggr) \biggr\} ,
\end{eqnarray*}
and $M$ indicates the $M$th iteration step in MCMC.

\subsection{Implementation}

The Gibbs sampling starts at~\ref{A-11}, with initial values for
$\theta^{(0)}$, $c^{(0)}$, $\phi^{(0)}$, $\varphi^{(0)}$,
$\eta^{(0)}$, $\delta^{(0)}$, $\tau^{(0)}$ and $\nu^{(0)}$, and then
loops through~\ref{A-12} until the MCMC has converged. The initial
values chosen in the applications were $\theta^{(0)}=\vec{0}$,
$c^{(0)}=\vec{0}$, $\phi^{(0)}=1$, $\varphi^{(0)}=\vec{0}$,
$\eta^{(0)}=\vec{0}$, $\delta^{(0)}=\vec{1}$, $\tau^{(0)}=\vec{1}$
and $\nu^{(0)}=\vec{1}$, where we used ``$\vec{a}$'' here to indicate
that each element of the corresponding vector or set has the same value ``$a$''.
The convergence was evaluated informally by
looking at trace plots, and was found to obtain at most after 30,000
of 50,000 iterations in the examples.

\section{Characteristics of 25 studied individuals}\label{appB}
Twenty-five individuals from the MetaMetrics data base are studied
in detail; the characteristics of the data for these individuals are
described in Table~\ref{characteristics}.
%
\begin{table}
\caption{Characteristics of the 25 considered individuals from the MetaMetrics
data}
\label{characteristics}
\begin{tabular*}{\textwidth}{@{\extracolsep{\fill}}ld{3.0}d{3.0}d{2.0}cd{3.0}c@{}}
\hline
& \multicolumn{1}{c}{\textbf{Total tests}} & \multicolumn{1}{c}{\textbf{Days}} &
\multicolumn{1}{c}{\textbf{Max. tests/days}} & \multicolumn{1}{c}{\textbf{Range of items/test}} & \multicolumn{1}{c}{\textbf{Max. gap}}
& \multicolumn{1}{c@{}}{\textbf{Grade}} \\
\hline
No.~1 & 147 & 73 & 8 & 4--25 & 105 & 4\\
No.~2 & 162 & 64 & 9 & 3--17 & 102 & 2\\
No.~3 & 118 & 77 & 4 & 3--21 & 87 & 2\\
No.~4 & 93 & 53 & 4 & 5--25 & 147 & 2\\
No.~5 & 114 & 89 & 3 & 6--25 & 109 & 2\\
No.~6 & 157 & 57 & 29 & 4--20 & 116 & 2\\
No.~7 & 153 & 63 & 7 & 4--20 & 97 & 2\\
No.~8 & 60 & 50 & 5 & 3--24 & 168 & 6\\
No.~9 & 135 & 53 & 7 & 4--24 & 93 & 2\\
No.~10 & 137 & 54 & 6 & 4--17 & 219 & 1\\
No.~11 & 214 & 100 & 11 & 3--18 & 108 & 2\\
No.~12 & 113 & 76 & 4 & 4--16 & 45 & 2\\
No.~13 & 95 & 65 & 4 & 4--14 & 113 & 2\\
No.~14 & 116 & 57 & 6 & 5--17 & 107 & 2\\
No.~15 & 155 & 71 & 9 & 4--20 & 107 & 1\\
No.~16 & 247 & 76 & 13 & 3--19 & 113 & 2\\
No.~17 & 254 & 76 & 12 & 3--18 & 107 & 2\\
No.~18 & 304 & 53 & 31 & 3--12 & 49 & 2\\
No.~19 & 167 & 83 & 5 & 3--23 & 58 & 2\\
No.~20 & 101 & 68 & 9 & 4--23 & 117 & 2\\
No.~21 & 88 & 58 & 9 & 3--23 & 110 & 2\\
No.~22 & 220 & 96 & 8 & 2--23 & 104 & 3\\
No.~23 & 80 & 66 & 6 & 2--25 & 93 & 6\\
No.~24 & 105 & 60 & 6 & 6--24 & 62 & 3\\
No.~25 & 218 & 74 & 12 & 3--25 & 113 & 2\\
\hline
\end{tabular*}
\end{table}

\section{Posterior propriety}\label{appC}
%
\begin{theorem}
Suppose $n\geq2$ and, for $i=1,\ldots,n$, $T_i\geq2$ and $S_{i,t}
\geq2$ for at least two days $t \in\{1, \ldots, T_i\}$ with at
least two of the tests on each of the two days having at least one 0
and one 1 observation. Then the posterior density of the DIR model
is proper.
\end{theorem}

We first give some needed lemmas that may be of independent interest
for proving posterior propriety in other logistic modeling
scenarios. Proofs of these lemmas are given in Appendix A of \citet{Wang}.

%
\begin{lemma}\label{lem-11}
For any three real numbers $x$, $\varepsilon_1$ and $\varepsilon_2$,
\[
\frac{e^{x+\varepsilon_1}}{1+e^{x+\varepsilon_1}}\times\frac{1}{1+e^{x+
\varepsilon_2}}\leq e^{-|x|+|\varepsilon_1|+|\varepsilon_2|} .
\]
\end{lemma}

%
\begin{lemma}\label{lem-12}
For $\theta_i \in(-\infty, \infty)$, $i=1,2$,
\begin{eqnarray*}
&&\int_{-\infty}^{\infty}\int_{-\infty}^{\infty}
\int_0^{\infty} \tau^{-1/2} e^{-\tau(\eta_1^2+\eta_2^2)}
e^{-(|\theta_1+\eta_1|+|\theta_1-\eta_1|+|\theta_2+\eta_2|+|\theta
_2-\eta_2|)}\,d\tau \,d\eta_1 \,d\eta_2\\
&&\qquad \leq K
e^{-(|\theta_1|+|\theta_2|)} ,
\end{eqnarray*}
with some constant $K$.
\end{lemma}

%
\begin{lemma}\label{lem-13}
For $\theta_i \in(-\infty, \infty)$, $i=1,2$,
\begin{eqnarray*}
\int_{-\infty}^{\infty} \int_{-\infty}^{\infty}
\int_{0}^{\infty} \delta^{-1/2}
e^{-({\delta}/{2})(\varphi_1^2+\varphi_2^2)}e^{-(|\theta_1+\varphi
_1|+|\theta_2+\varphi_2|)} \,d\delta \,d\varphi_1 \,d
\varphi_2 \leq\frac{K}{1+ |\theta_1|} , 
\end{eqnarray*}
with some constant $K$.
\end{lemma}

%
\begin{lemma}\label{lem-14}
For $T\geq2$,
\begin{eqnarray*}
&&\hspace*{-4pt}\int_0^{\infty}\int_0^{\infty}
\int_0^{\infty}\int_{-\infty}^{\infty}
\int_{-\infty}^{\infty} \frac{1}{\phi^{3/2}} \cdot
\frac{1}{1+|\sqrt{{B(c)}/{\phi}}z+A(c)|}e^{-{z^2}/{2}}
\\
&&\hspace*{70pt}\qquad{}\times  \frac{1}{1+|\sqrt{B'(c')/\phi}z' +A'(c')|}e^{-{z'^2}/{2}} \,d z\, d z'\,d c\, d
c' \,d\phi<\infty,
\end{eqnarray*}
where
\begin{eqnarray*}
A(c)&=&\mu_{G_j}\prod_{t=1}^T
\bigl(1-c\rho\Delta_t^+\bigr)+\sum_{t=1}^T
c\Delta_t^+\prod_{i=t+1}^T
\bigl(1-c\rho\Delta_i^+\bigr),
\\
B(c)&=&\sum_{t=1}^T \Delta_t
\prod_{i=t+1}^T\bigl(1-c\rho
\Delta_i^+\bigr)^2+\phi V_{G_j} \prod
_{t=1}^T \bigl(1-c\rho\Delta_{t}^+
\bigr)^2,
\\
A'\bigl(c'\bigr)&=&\mu_{G_j}\prod
_{t=1}^T\bigl(1-c'\rho
\Delta_t^+\bigr)+\sum_{t=1}^T
c'\Delta_t^+\prod_{i=t+1}^T
\bigl(1-c'\rho\Delta_i^+\bigr),
\\
B'\bigl(c'\bigr)&=&\sum_{t=1}^T
\Delta_t \prod_{i=t+1}^T
\bigl(1-c'\rho\Delta_i^+\bigr)^2+\phi
V_{G_j} \prod_{t=1}^T
\bigl(1-c'\rho\Delta_{t}^+\bigr)^2 ,
\end{eqnarray*}
and we have dropped the label $i$ in the subscripts for $\Delta_{i,t}$,
$c_i$, $\mu_{G_{j_i}}$ and $V_{G_{j_i}}$.
\end{lemma}

%
\begin{lemma}\label{lem-15}
For $T\geq2$,
\[
\int_{0}^{\infty}\int_{-\infty}^{\infty}
\frac{1}{1+|\sqrt{{B(c)}/{\phi}}z+A(c)|}\exp\biggl\{-\frac
{z^2}{2}\biggr\} \,d z \,dc <\infty,
\]
with $A(c)$ and $B(c)$ defined in Lemma~\ref{lem-14}.
\end{lemma}

\begin{pf}
In proving posterior propriety, it is easiest to work with the
posterior density without the data augmentation, namely,
%
\begin{eqnarray}\label{eqn-119}
&& \pi(\theta,c,\tau, \eta, \varepsilon, \phi\mid X)
\nonumber\hspace*{-32pt}\\
&&\quad\propto \Biggl\{\prod_{i=1}^n
\frac{1}{\sqrt{2\pi}V_{G_{j_i}}}\exp\biggl(-\frac{(\theta_{i,0}-\mu
_{G_{j_i}})^2}{2
V_{G_{j_i}}} \biggr)\mathbf{I}_{\{c_i\geq0\}}
\frac{1}{\tau_i^{3/2}}\frac{1}{\delta_i^{3/2}} \Biggr\} \frac{1}{\phi^{3/2}}
\nonumber\hspace*{-32pt}
\\
&&\qquad\times \Biggl\{\prod_{i=1}^n \prod
_{t=1}^{T_i} \prod
_{s=1}^{S_{i,t}} \prod_{l=1}^{K_{i,t,s}}
\frac{1}{\sqrt{2\pi}\sigma}\exp\biggl(-\frac{\varepsilon_{i,t,s,l}^2} {
2\sigma^2} \biggr) \Biggr\} \Biggl\{\prod
_{i=1}^n\prod_{t=1}^{T_i}
\sqrt{\frac{\delta_i}{2\pi}}\exp\biggl(-\frac{\delta_i\varphi
_{i,t}^2}{2} \biggr) \Biggr\}
\nonumber\hspace*{-32pt}
\\
&&\qquad\times  \Biggl\{\prod_{i=1}^n \prod
_{t=1}^{T_i} \biggl(\frac{\tau_i}{2\pi}
\biggr)^{{(S_{i,t}-1)}/{2}}\exp{ \biggl(-\frac{\tau_i
\eta_{i,t}^{*\prime}\Sigma_{i,t}^{-1}\eta_{i,t}^*}{2} \biggr)} 
\Biggr
\}\hspace*{-32pt}
\\
&&\qquad\times  \Biggl\{\prod_{i=1}^n \prod
_{t=1}^{T_i} \prod
_{s=1}^{S_{i,t}}\prod_{l=1}^{K_{i,t,s}}
\frac{\exp
[X_{i,t,s,l}(\theta_{i,t}-a_{i,t,s}+\varphi_{i,t}+\eta_{i,t,s}+\varepsilon
_{i,t,s,l}) ]} {
1+\exp(\theta_{i,t}-a_{i,t,s}+\varphi_{i,t}+\eta_{i,t,s}+\varepsilon
_{i,t,s,l})}\nonumber\hspace*{-32pt}\\
&&\hspace*{196pt}{}\times I\Biggl\{\eta_{i,t,S_{i,t}}=-\sum_{s=1}^{S_{i,t}-1}
\eta_{i,t,s}\Biggr\} \Biggr\}
\nonumber\hspace*{-32pt}
\\
&&\qquad\times  \Biggl\{\prod_{i=1}^n \prod
_{t=1}^{T_i} \sqrt\frac{\phi}{2\pi\Delta_{i,
t}}\exp
\biggl(-\frac{\phi\{\theta_{i,t}-\theta_{i,t-1}-c_i(1-\rho\theta_{i,t-1})
\Delta_{i,t}^+\}^2}{2\Delta_{i,
t}} \biggr) \Biggr\} .\hspace*{-32pt}
\nonumber
\end{eqnarray}
Noting that
\[
\frac{\exp[X_{i,
t,s,l}(\theta_{i,t}-a_{i,t,s}+\varphi_{i,t}+\eta_{i,t,s}+\varepsilon
_{i,t,s,l}) ]} {
1+\exp(\theta_{i,t}-a_{i,t,s}+\varphi_{i,t}+\eta_{i,t,s}+\varepsilon
_{i,t,s,l})} \leq1 ,
\]
an upper bound on the posterior density can be found by dropping all
terms except the 0 and 1 test observations
in the assumed tests for each individual.
Utilizing
Lemma~\ref{lem-11} for each pair of observations 0 and 1 then results
in the following upper bound on the posterior density
(\ref{eqn-119}):
%
\begin{eqnarray}\label{eqn-a11}
&& \frac{1}{\phi^{3/2}} \Biggl\{\prod_{i=1}^n
\frac{1}{\sqrt{2\pi}V_{G_{j_i}}}\exp\biggl(-\frac{(\theta_{i,0}-\mu
_{G_{j_i}})^2}{2
V_{G_{j_i}}} \biggr)\mathbf{I}_{\{c_i\geq
0\}}
\frac{1}{\tau_i^{3/2}}\frac{1}{\delta_i^{3/2}} \Biggr\}
\nonumber\hspace*{-32pt}
\\
&&\quad{}\times  \Biggl\{\prod_{i=1}^n\prod
_{t=1}^{T_i} \prod
_{s=1}^{S_{i,t}}\prod_{\ell=1}^{K_{i,t,s}}
\frac{1}{\sqrt{2\pi}\sigma}\exp\biggl(-\frac{\varepsilon_{i,t,s,l}^2} {
2\sigma^2} \biggr) \Biggr\} \Biggl\{\prod
_{i=1}^n \prod
_{t=1}^{T_i}\sqrt{\frac{\delta_i}{2\pi}}\exp\biggl(-
\frac{\delta_i\varphi_{i,t}^2}{2} \biggr) \Biggr\}
\nonumber\hspace*{-32pt}
\\
&&\quad{}\times \Biggl\{\prod_{i=1}^n\prod
_{t=1}^{T_i}\biggl(\frac{\tau_i} {
2\pi}
\biggr)^{{(S_{i,t}-1)}/{2}}\exp{ \biggl(-\frac{\tau_i{\eta_{i,t}^{*\prime}}
\Sigma_{i,t}^{-1}\eta_{i,t}^*}{2} \biggr)} 
\Biggr
\}\nonumber\hspace*{-32pt}\\
&&\quad{}\times \Biggl\{\prod_{i=1}^n \prod
_{t=1}^{T_i} \prod_{s=1}^{S_{i,t}}I
\Biggl\{\eta_{i,t,S_{i,t}}=-\sum_{s=1}^{S_{i,t}-1}
\eta_{i,t,s}\Biggr\} \Biggr\}
\nonumber\hspace*{-32pt}
\\
&&\quad{}\times \Biggl\{\prod_{i=1}^n \exp\bigl( -|
\theta_{i,t_i}+\varphi_{i,t_i}+\eta_{i,t_i,m}|+|a_{i,t_i,m}|+
|\varepsilon_{i,t_i,m,k}|+|\varepsilon_{i,t_i,m,k'}| \bigr)\hspace*{-32pt}
\\
&&\quad{}\times \exp\bigl( -|\theta_{i,t_i}+\varphi_{i,t_i}+
\eta_{i,t_i,m'}|+|a_{i,t_i,m'}|+ |\varepsilon_{i,t_i,m',h}|+|
\varepsilon_{i,t_i,m',h'}| \bigr)
\nonumber\hspace*{-32pt}\\
&&\quad{}\times\exp\bigl( -|\theta_{i,t_i'}+\varphi_{i,t_i'}+
\eta_{i,t_i',r}|+|a_{i,t_i',r}|+ |\varepsilon_{i,t_i',r,q}|+|
\varepsilon_{i,t_i',r,q'}| \bigr)
\nonumber\hspace*{-32pt}
\\
&&\quad{}\times\exp\bigl( -|\theta_{i,t_i'}+\varphi_{i,t_i'}+
\eta_{i,t_i',r'}|+|a_{i,t_i',r}|+ |\varepsilon_{i,t_i',r',g}|+|
\varepsilon_{i,t_i',r',g'}| \bigr)\Biggr\}
\nonumber\hspace*{-40pt}
\\
&&\quad{}\times \Biggl\{\prod_{i=1}^n \prod
_{t=1}^{T_i} \sqrt\frac{\phi}{2\pi\Delta_{i,
t}}\exp
\biggl(-\frac{\phi\{\theta_{i,t}-\theta_{i,t-1}-c_i(1-\rho\theta
_{i,t-1})\Delta_{i,t}^+\}^2}{2\Delta_{
i,t}} \biggr) \Biggr\} .
\nonumber\hspace*{-32pt}
\end{eqnarray}
Ignoring multiplicative constants, and integrating out all the
$\varepsilon_{i,t,s,l}$, (\ref{eqn-a11}) has an upper bound of
%
\begin{eqnarray}\label{eqn-13}
& & \frac{1}{\phi^{3/2}} \Biggl\{\prod_{i=1}^n
\frac{1}{\sqrt{2\pi}V_{G_{j_i}}}\exp\biggl(-\frac{(\theta_{i,0}-\mu
_{G_{j_i}})^2}{2
V_{G_{j_i}}} \biggr)\mathbf{I}_{\{c_i\geq
0\}}
\frac{1}{\tau_i^{3/2}}\frac{1}{\delta_i^{3/2}} \Biggr\}
\nonumber
\\
&&\qquad{}\times \Biggl\{\prod_{i=1}^n \prod
_{t=1}^{T_i}\sqrt{\frac{\delta_i}{2\pi}}\exp
\biggl(-\frac{\delta_i\varphi_{i,t}^2}{2} \biggr) \Biggr\}\nonumber\\
&&\qquad{}\times \Biggl\{\prod
_{i=1}^n\prod_{t=1}^{T_i}
\biggl(\frac{\tau_i} {
2\pi}\biggr)^{{(S_{i,t}-1)}/{2}}\exp{ \biggl(-\frac{\tau_i{\eta_{i,t}^{*\prime}}
\Sigma_{i,t}^{-1}\eta_{i,t}^*}{2}
\biggr)} \Biggr\} 
\\
&&\qquad{}\times \Biggl\{\prod_{i=1}^n \prod
_{t=1}^{T_i} \prod
_{s=1}^{S_{i,t}}I\Biggl\{\eta_{i,t,S_{i,t}}=-\sum
_{s=1}^{S_{i,t}-1}\eta_{i,t,s}\Biggr\} \Biggr\}
\nonumber
\\
&&\qquad{}\times \Biggl\{\prod_{i=1}^n \exp\bigl\{ -|
\theta_{i,t_i}+\varphi_{i,t_i}+\eta_{i,t_i,m}|\bigr\}\exp\bigl\{ -|
\theta_{i,t_i}+\varphi_{i,t_i}+\eta_{i,t_i,m'}|\bigr\}
\nonumber
\\
&&\hspace*{32pt}\qquad{}\times\exp\bigl\{ -|\theta_{i,t_i'}+\varphi_{i,t_i'}+
\eta_{i,t_i',r}|\bigr\}\exp\bigl\{ -|\theta_{i,t_{i'}}+\varphi_{i,t_{i'}}+
\eta_{i,t_i',r'}|\bigr\} \Biggr\}
\nonumber
\\
&&\qquad{}\times \Biggl\{\prod_{i=1}^n\prod
_{t=1}^{T} \sqrt\frac{\phi}{2\pi\Delta_{
i,t}}\exp
\biggl(-\frac{\phi\{\theta_{i,t}-\theta_{i,t-1}-c_i(1-\rho\theta
_{i,t-1})\Delta_{i,t}^+\}^2}{2\Delta_{
i,t}} \biggr) \Biggr\} .
\nonumber
\end{eqnarray}
%

We only consider here the ``least information'' case in which
\mbox{$S_{i,t_i}=S_{i,t_{i'}}=2$}; the more general case can be done
similarly. Then $\eta_{i,t_i,m} = -\eta_{i,t_i,m'}$,
$\eta_{i,t_{i'},r}=-\eta_{i,t_{i'},r'}$,
$\exp{ (-\tau_i{\eta_{i,t_i}^{*\prime}}
\Sigma_{i,t_i}^{-1}\eta_{i,t_i}^*/2 )} = \exp{ (-\tau_i
\eta_{i,t_i,m}^2 )}$, and
$\exp (-\tau_i{\eta_{i,t_{i'}}^{*\prime}}
\times \break\Sigma_{i,t_{i'}}^{-1}\eta_{i,t_{i'}}^*/2 ) =
\exp{ (-\tau_i \eta_{i,t_{i'},r}^2 )}$. Using this in
(\ref{eqn-13}) and integrating out all other $\eta$ except for
$\eta_{i,t_i,m}$ and $\eta_{i,t_{i'},r}$ and all $\varphi$ except
for $\varphi_{i,t_i}$ and $\varphi_{i,t_i'}$, results in the
expression
\begin{eqnarray*}
& & \frac{1}{\phi^{3/2}} \Biggl\{\prod_{i=1}^n
\frac{1}{\sqrt{2\pi}V_{G_{j_i}}}\exp\biggl(-\frac{(\theta_{i,0}-\mu
_{G_{j_i}})^2}{2
V_{G_{j_i}}} \biggr) \mathbf{I}_{\{c_i\geq
0\}}
\Biggr\}
\nonumber
\\
&&\qquad{}\times \Biggl\{\prod_{i=1}^n
\frac{1}{\delta_i^{3/2}}\frac{\delta_i}{2\pi}\exp\biggl(-\frac{\delta
_i\varphi_{i,t_i}^2}{2} \biggr) \exp
\biggl(-\frac{\delta_i\varphi_{i,t_i'}^2}{2} \biggr) \cdot\frac{1}{\tau
_i^{3/2}} \\
&&\hspace*{34pt}\qquad{}\times{
\frac{\tau_i}{2\pi}}\exp{ \bigl(-\tau_i\bigl(\eta_{i,t_i,m}^2
+\eta_{i,t_i',r}^2\bigr) \bigr)}
\nonumber
\\
&&\hspace*{34pt}\qquad{}\times \exp\bigl\{-\bigl(|\theta_{i,t_i}+\varphi_{i,t_i}+
\eta_{i,t_i,m}| +|\theta_{i,t_i}+\varphi_{i,t_i}-
\eta_{i,t_i,m}|\bigr)\bigr\}
\nonumber
\\
&&\hspace*{37pt}\qquad{}\times \exp\bigl\{-\bigl(|\theta_{i,t_i'}+\varphi_{i,t_i'}+
\eta_{i,t_i',r}| +|\theta_{i,t_i'}+\varphi_{i,t_i'}-
\eta_{i,t_i',r}|\bigr)\bigr\} \Biggr\}
\nonumber
\\
&&\qquad{}\times \Biggl\{\prod_{i=1}^n \prod
_{t=1}^{T_i} \sqrt\frac{\phi}{2\pi\Delta_{i,
t}}\exp
\biggl(-\frac{\phi\{\theta_{i,t}-\theta_{i,t-1}-c_i(1-\rho\theta
_{i,t-1})\Delta_{i,t}^+\}^2}{2\Delta_{i,
t}} \biggr) \Biggr\} .
\end{eqnarray*}
Next integrate out over $\tau_i$, $\eta_{i,t_i,m}$ and $\eta
_{i,t_i',r}$ using Lemma~\ref{lem-12},
resulting in the upper bound (again ignoring multiplicative
constants)
%
\begin{eqnarray}\label{eqn-14}
& & \frac{1}{\phi^{3/2}} \Biggl\{\prod_{i=1}^n
\frac{1}{\sqrt{2\pi}V_{G_{j_i}}}\exp\biggl(-\frac{(\theta_{i,0}-\mu
_{G_{j_i}})^2}{2
V_{G_{j_i}}} \biggr) \mathbf{I}_{\{c_i\geq
0\}}
\Biggr\}
\nonumber
\\
&&\qquad{}\times \Biggl\{\prod_{i=1}^n
\frac{1}{\delta_i^{3/2}}\frac{\delta_i}{2\pi}\exp\biggl(-\frac{\delta
_i\varphi_{i,t_i}^2}{2} \biggr) \exp
\biggl(-\frac{\delta_i\varphi_{i,t_i'}^2}{2} \biggr)
\\
&&\hspace*{54pt}{}\times  \exp\bigl\{
-\bigl(|\theta_{i,t_i}+
\varphi_{i,t_i}|+|\theta_{i,t_i'}+\varphi_{i,t_i'}|\bigr)\bigr\}
\Biggr\}
\nonumber\\
&&\qquad{}\times\Biggl\{\prod_{i=1}^n \prod
_{t=1}^{T_i} \sqrt\frac{\phi}{2\pi\Delta_{i,
t}}\exp
\biggl(-\frac{\phi\{\theta_{i,t}-\theta_{i,t-1}-c_i(1-\rho\theta
_{i,t-1})\Delta_{i,t}^+\}^2}{2\Delta_{i,
t}} \biggr) \Biggr\} .
\nonumber
\end{eqnarray}
Next integrate out $\delta_i$, $\varphi_{i,t_i}$ and
$\varphi_{i,t_i'}$ using Lemma~\ref{lem-13}. The resulting upper bound on~(\ref{eqn-14}) is
\begin{eqnarray*}\label{eqn-18}
& & \frac{1}{\phi^{3/2}} \Biggl\{\prod_{i=1}^n
\frac{1}{\sqrt{2\pi}V_{G_{j_i}}}\exp\biggl(-\frac{(\theta_{i,0}-\mu
_{G_{j_i}})^2}{2
V_{G_{j_i}}} \biggr)\mathbf{I}_{\{c_i\geq0\}}
\cdot\frac{1}{1+|\theta_{i,t_i'}|} \Biggr\}
\\
&&{}\qquad\times \Biggl\{\prod_{i=1}^n\prod
_{t=1}^{T_i} \sqrt\frac{\phi}{2\pi\Delta_{
i,t}}\exp
\biggl(-\frac{\phi\{\theta_{i,t}-\theta_{i,t-1}-c_i(1-\rho\theta
_{i,t-1})\Delta_{i,t}^+\}^2}{2\Delta_{
i,t}} \biggr) \Biggr\} .
\end{eqnarray*}
Integrating out all the $\theta_{i,t}$ except the $\theta_{i,t_i'}$
results in the expression
%
\begin{eqnarray}\label{eqn-112}
&& \frac{1}{\phi^{3/2}} \Biggl\{\prod_{i=1}^n
\mathbf{I}_{\{c_i\geq0\}} \cdot\frac{1}{1+|\theta_{i,t_i'}|} \Biggr\}
\nonumber
\\
&&{}\qquad\times \Biggl\{\prod_{i=1}^n \sqrt{
\frac{\phi}{2\pi V_{G_{j_i}}}}\nonumber\\
&&\hspace*{34pt}\qquad{}\times \Biggl({1}\bigg/\Biggl(\sum_{t=1}^{t_i'} \Delta_{i,t}
\prod_{i=t+1}^{t_i'}\bigl(1-c_i\rho\Delta_{i,t}^+\bigr)^2
\nonumber
\\[-8pt]
\\[-8pt]
\nonumber
&&\hspace*{136pt}\qquad{}+\phi V_{G_{j_i}} \prod_{t=1}^{t_i'} \bigl(1-c_i\rho\Delta_{i,t}^+\bigr)^2\Biggr) \Biggr)^{1/2}
\\
&&{}\qquad\times \exp\Biggl(-\Biggl(\phi\Biggl(\theta_{i,t_i'}-\mu_{G_{j_i}}\prod
_{t=1}^{t_i'}\bigl(1-c_i\rho\Delta_{i,t}^+\bigr)\nonumber\\
&&\hspace*{53pt}\qquad\quad{}-\sum_{t=1}^{t_i'}
c_i\Delta_{i,t}^+\prod_{i=t+1}^{t_i'}\bigl(1-c_i\rho
\Delta_{i,t}^+\bigr)\Biggr)^2\Biggr)\nonumber\\
&&\qquad\quad{}\bigg/\Biggl(2\Biggl(\sum_{t=1}^{t_i'}
\Delta_{i,t}\prod_{i=t+1}^{t_i'}\bigl(1-c_i\rho\Delta_{i,t}^+\bigr)^2+\phi
V_{G_{j_i}} \prod_{t=1}^{t_i'}
\bigl(1-c_i\rho\Delta_{i,t}^+\bigr)^2\Biggr)\Biggr) \Biggr) \Biggr\} .
\nonumber
\end{eqnarray}
Finally, defining
\[
z_i = \frac{\sqrt{\phi}(\theta_{i,t_i'} - A_i(c_i))}{\sqrt{B_i(c_i)}} ,
\]
using Lemma~\ref{lem-15} to integrate out all $\theta_{i,t_i'}$ and
$c_i$, except for two individuals, and then using Lemma
\ref{lem-14} for the remaining variables of (\ref{eqn-112}), it
follows that the integral is finite, completing the proof.
\end{pf}
\end{appendix}

\section*{Acknowledgments}
The research of Xiaojing Wang was part of her dissertation at
the Department of Statistical Science, Duke University.
The authors are grateful to Jack Stenner, Hal Burdick, Carl
Swartz and Sean Hanlon at MetaMetrics Inc. for valuable
discussions, and to the Editor, Associate Editor and referees
for numerous suggestions that significantly improved the paper.

\bibliographystyle{imsart-nameyear}

%


\printaddresses

\end{document}